\RequirePackage{fix-cm}
\documentclass[twocolumn]{svjour3}          
\smartqed  
\usepackage{graphicx}
\usepackage{amsmath}
\usepackage{siunitx}
\usepackage{amssymb}
\usepackage{xcolor}

\newcommand{\be}{\begin{equation}}
\newcommand{\ee}{\end{equation}}
\newcommand{\bee}{\begin{eqnarray}}
\newcommand{\eee}{\end{eqnarray}}

\begin{document}

\title{Rate-dependent drag instability in granular materials}



\author{T. Hossain \and P. Rognon
}

\authorrunning{T. Hossain \& al.} 

\institute{T. Hossain and   P.  Rognon \at
             School of Civil Engineering, The University of Sydney, Sydney,
2006 NSW, Australia. \\
              \email{pierre.rognon@sydney.edu.au}}

\date{Received: 15/11/2019;  Revised: 9/04/2020; Accepted: Under further review by Granular Matter}

\maketitle

\begin{abstract}
We investigate the conditions leading to large drag force fluctuations in granular materials. The study is based on a set of experimental drag tests, which involve pulling a plate vertically through a cohesionless granular material. In agreement with previous observations, drag force exhibits significant and sudden drops -up to 60\%- when the plate is pulled out at low velocities. We further find that this instability vanishes at higher pullout velocities and near the surface. We empirically characterise the frequency and amplitude of these fluctuations and find that these properties are not consistent with a classical stick-slip dynamics. We therefore propose an alternative physical mechanism that can explain these force fluctuations.
\end{abstract}

\section{Introduction}

Objects moving through Newtonian liquids experience a drag force hindering their motion. Stokes drag and turbulent drag models capture this force at low and high Reynolds number, respectively.  Similarly, objects moving through a packing of grains are subjected to a drag force. However, granular drag forces are fundamentally different from Stokes and turbulent drags. 

At low velocities, granular drag is rate-independent and proportional to pressure. It is thus referred to as \textit{frictional} drag. Several studies have evidenced such a frictional drag using objects with different geometries, being moved vertically or horizontally, and embedded at different depths \cite{albert1999slow,albert2000jamming,albert2001granular,miller1996stress,nguyen1999properties,gravish2010force,costantino2011low,ding2011drag,guillard2013depth}.  At higher velocities, granular drag becomes rate-dependent. Specifically, it increases quadratically with the velocity, which is reminiscent of a turbulent drag \cite{percier2011lift,potiguar2013lift,takehara2010high,takehara2014high}. The underlying mechanisms are (i) the shear stress associated with shear deformation within the granular packing  and (ii) the inertia associated with moving grain around the object, which are both necessary to enable its motion.

Granular drag forces are not always steady. For instance, a drag instability was evidenced in ploughing experiments involving moving a vertical blade horizontally through cohesionless grains \cite{gravish2010force}. Large fluctuations in drag force were observed when the granular packing was initially dense enough. The underlying mechanism involves cycles in deformation of the granular packing surface. As the blade moves forward, a hump grows in front of it, which coincide with an increase in drag force. Periodically, an avalanche at the hump surface is triggered that temporarily relaxes the drag force.

A similar instability in drag force was observed in a different experimental configuration, involving uplifting vertically a horizontal plate through semolina \cite{duri2017vertical}. Even at a depth where the plate motion did not affect the geometry of the free surface,  large fluctuations in drag force were observed when pulling the plate at a constant velocity. These fluctuations decreased and vanished when the plate moved closer to the packing surface. However, the origin of this instability is poorly understood. As a result, the conditions leading to its occurrence are not known. \\

The purpose of this paper is to establish the conditions leading to granular drag instability during uplift. To this aim, we conducted a series of laboratory experiments involving driving a plate at a constant velocity and measuring the evolution of the drag force. Tests covered a range of uplift velocities and grain size. The paper is organised as follows. In Section \ref{sec:method}, we present the experimental method used to measure drag force and its fluctuations. In Section \ref{sec:empirical}, we empirically characterise the drag fluctuations by quantifying their typical frequency and magnitude. In Section \ref{sec:discusion}, we introduce a physical process that can explain the observed properties of this drag instability.


\section{Experimental method} \label{sec:method} 

This Section presents the experimental method used to conduct vertical uplift tests and measure the resulting drag forces. 

\subsection{Materials and set-up}

\begin{figure}[htb!]
\includegraphics[width=1\columnwidth,trim={1.5cm 7cm 6cm 0cm}, clip]{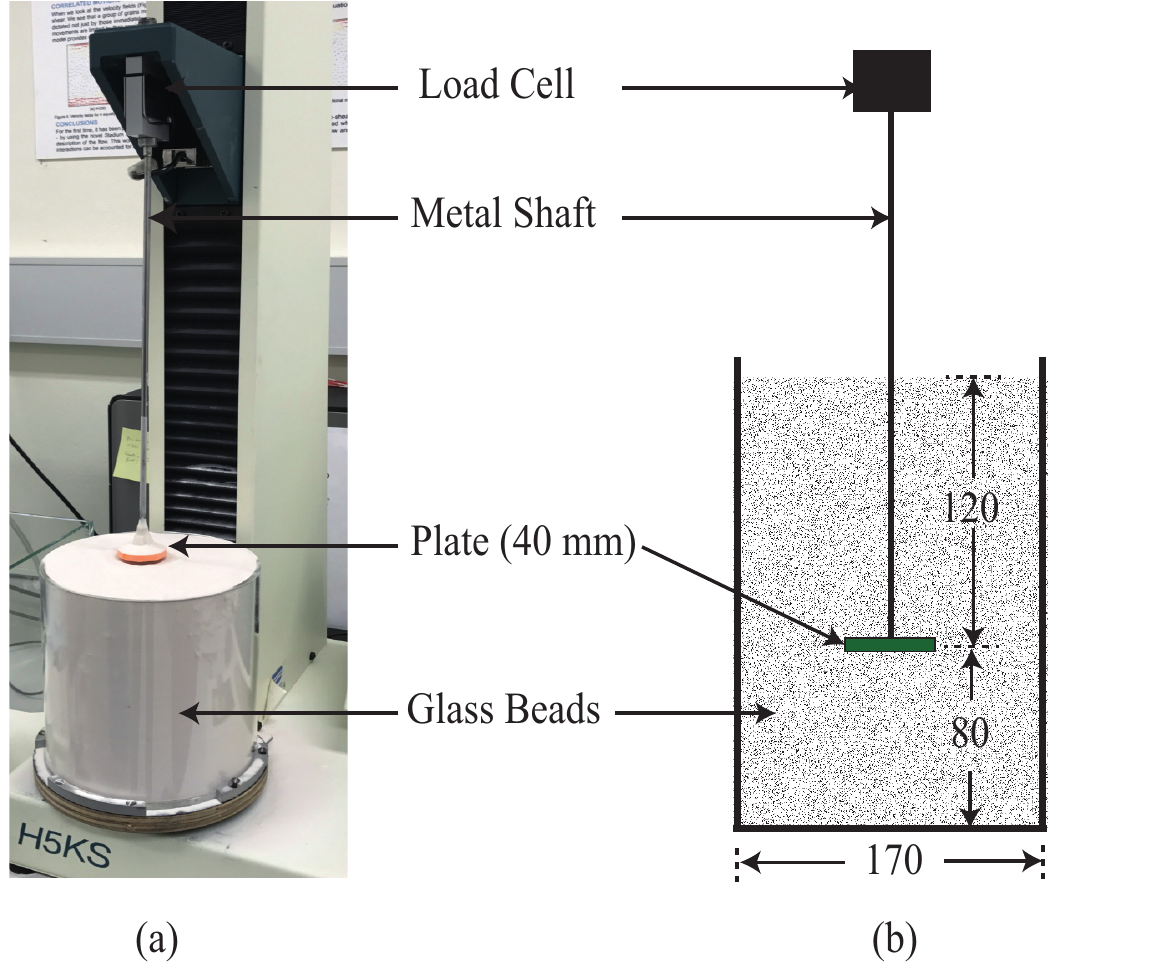}
\caption{Experimental method. (a) Photo of the experimental set-up taken at the end of a test. (b) Illustration of the initial position of the plate in the glass bead container (distances are expressed in milimeters).}
\label{exp_setup}       
\end{figure}

Uplift tests were conducted using the experimental set-up presented in figure \ref{exp_setup}. This set-up has previously been used in \cite{dyson2014pull} to measure the uplift capacity of anchors with a shape mimicking tree roots. It is comprised of a container filled with glass beads, in which a PDMS disk-shape plate is buried at a designated depth. The plate's motion is driven by a loading frame (H5KS Olsen Loading Frame) via a stainless steel shaft. 

The plate has a circular cross section of diameter $B= \SI{40}{mm}$, and is $\SI{4}{mm}$ thick. 
The container diameter is \SI{170}{mm}, which is about four times larger than the plate. Three categories of glass beads are used, with an average grain diameter of either $ d =$ \SI{50}{\micro\meter}, \SI{300}{\micro\meter} or \SI{1}{ mm}. For each category, the grain size distribution is $d\pm10\%$. The advantage of using spherical glass beads as opposed to angular grains such as sand is that they consistently form a dense packing with no critical dependence on the mode of pouring. This favours test reproducibility, as discussed below. These packings are characterised by an internal friction angle  $\phi\approx$ \SI{23}{^o}, a specific weight \SI{23.7}{kN/m^3} - which is the product of the gravity acceleration and packing density- and a porosity of about \SI{20}{\%}.

\subsection{Uplift test procedure}

The preparation of the uplift tests involves filling the first \SI{80}{mm} of the container with glass beads, placing the plate horizontally at this location, and adding glass beads above it up to the desired level. All tests presented in the following  are performed with an initial plate depth of $H$=\SI{120}{mm} corresponding to an embedment ratio $\frac{H}{B} = 3$. Unless otherwise specified, the plate is placed at an equal distance from the vertical edge of the container (see figure \ref{exp_setup}b).

After this preparation, the uplift test involves vertically pulling the plate at a prescribed constant velocity $v$. The loading frame drives this motion using a DC servo motor and records over time the plate's vertical displacement $P(t)=vt$ as well as the drag force $F$ required to achieve it. The frequency of force and displacement readings is set to \SI{100}{Hz}. The force is zeroed during the preparation just before the plate is covered by grains. This means $F$ measures the net reaction of the granular packing on the plate and excludes the self-weight of the plate and shaft.  

We assessed the test reproducibility by measuring the uplift capacity (maximum value of the drag force $F(P)$) on tests repeated with the same conditions over the course of serval weeks. We found that the variability in uplift capacity was smaller than $5\%$. This suggests that the mode of preparation is reproducible and that potential variations in environmental factors such as air humidity do not critically influence the results. 

We also assessed the potential influence of the container finite size by repeating tests placing the plate closer and closer to an edge. Results showed no influence on the uplift capacity provided that the plate is not very close to the contained edge - a decrease in uplift capacity was observed for distances lesser than about \SI{10}{mm}. This suggests that tests conducted placing the plate in the centre of the container, which corresponds to a plate-to-container distance of \SI{65}{mm}, are not significantly influenced by the size of the container.

\subsection{Dimensional analysis}

Uplift tests involve several physical parameters including grain size $d$, grain density $\rho_g$, plate diameter $D$, plate depth $H$, gravity $g$, and uplift velocity $v$. From these, we define a set of characteristic times and dimensionless numbers that compare the rate at which the granular packing is being disturbed by the plate's motion and the rate at which it can recover from this disturbance. The most elementary time scale 
corresponds to the time $t_v$ for the plate to move up on a distance of one grain:

\be
 t_v = \frac{d}{v}.
\ee

\noindent This provides a basic time scale for the packing disturbance.

\subsubsection{Gravity relaxation time}

We define a first elementary relaxation time $t_g$ as:

\be
t_g = \sqrt{\frac{d}{g}},
\ee

\noindent which is proportional to the time for a grain to free fall on a distance $d$ under the action of the gravity. 
This relaxation time was found to be pivotal to explain the mobility of plates subjected to cyclic or transient loadings \cite{athani2018mobility,athani2019inertial}.
Accordingly, we introduce a dimensionless number $G$, which we refer to as \textit{relaxation} number, as the ratio of these two times:

\be
G = \frac{t_g}{t_v} = \frac{v}{\sqrt{gd}} 
\ee

\noindent Small values of $G$ means that grains' free-fall rearrangement is much faster than the packing disturbance induced by the plate motion; and vice-versa. Experiments were conducted using grain size ranging from \SI{0.05}{mm} to \SI{1}{mm} at velocities ranging from \SI{ 1.6}{ \times 10^{-5}  m/s} to \SI{1.6}{\times10^{-2}m/s}. This leads to values of $t_v$ ranging from \SI{3}{\times 10^{-3}s}  to  \SI{62}{s}, and values of $t_g$ ranging from \SI{2.2}{\times 10^{-3}s} to  \SI{}{10^{-2}s}. This translates into a value of $G$ ranging from:

\be
 3 \times 10^{-5} \lesssim G \lesssim 3
\ee

\noindent We note that $G$ is similar to a Froude number, which is often used in the context of granular flows on slopes to quantify the average speed $v$ of the flow \cite{pouliquen1999scaling}. 
We chose to use a different name for this dimensionless number to reflect the specificity of our experimental configuration.

\subsubsection{Shear and inertial times}

The plate's motion through the granular packing is expected to induce some shear deformation around it \cite{candelier2009creep,harich2011intruder,kolb2013rigid,seguin2019hysteresis}. While the precise extent of the zone being sheared is not known a priori in our configuration, it is expected to scale like the plate size $B$. Accordingly, we define a characteristic shear rate $v/B$, with a dimension of inverse time, and a typical inertial number $I$ as \cite{midi2004dense,da2005rheophysics}:

\be
I = \frac{v}{B}t_i; t_i = d\sqrt{\rho_g/\sigma_H}.
\ee

\noindent $I$ compares the shear time to the inertial time $t_i$. The inertial time measures the typical time for a grain of mass $m$ to move on a distance $d$ under the action of a force $\sigma_H d^2$, where $\sigma_H$ is the pressure in the packing at depth $H$: $\sigma_H \approx \phi \rho_g g H$. The corresponding inertial number is $I = \frac{v}{B} d \sqrt{\frac{1}{\phi gH}}$. The range of grain size and uplift velocity considered in our tests yields the following range for the inertial number:

 \be
 2 \times 10^{-8} \lesssim I \lesssim 4\times10^{-4}
\ee

\noindent This means that grains can rearrange inertially much faster than they are being sheared in all tests. This corresponds to a quasi-static shear regime, where the effective coefficient of friction of the granular packing is expected to be rate-independent \cite{midi2004dense}. Performing tests at higher inertial number could induce an increase in the effective friction coefficient of the packing. This effect could lead to a linear increase of the drag force with the plate velocity, as observed in Ref. \cite{hilton2013drag}.

\begin{table}[htp]
\begin{center}
\begin{tabular}{c|c|c|c|}
 $d$ [mm]& $D$ [mm] & $B = H/D$ & $v$ [ mm/s]\\
\hline
 $0.05  \rightarrow 1$ & 40& 3 &  \ $0.016  \rightarrow 16$ \\
\end{tabular}
\caption{Parameters of the uplift tests including grains size $d$, intruder width $D$, intruder embedment ratio $B$ and prescribed pullout velocity $v$.\label{tab:parameter}}
\end{center}
\end{table}%

\begin{figure}[h!]
 {\includegraphics[width=1\columnwidth,trim={0cm 0cm 0cm 0cm}, clip]{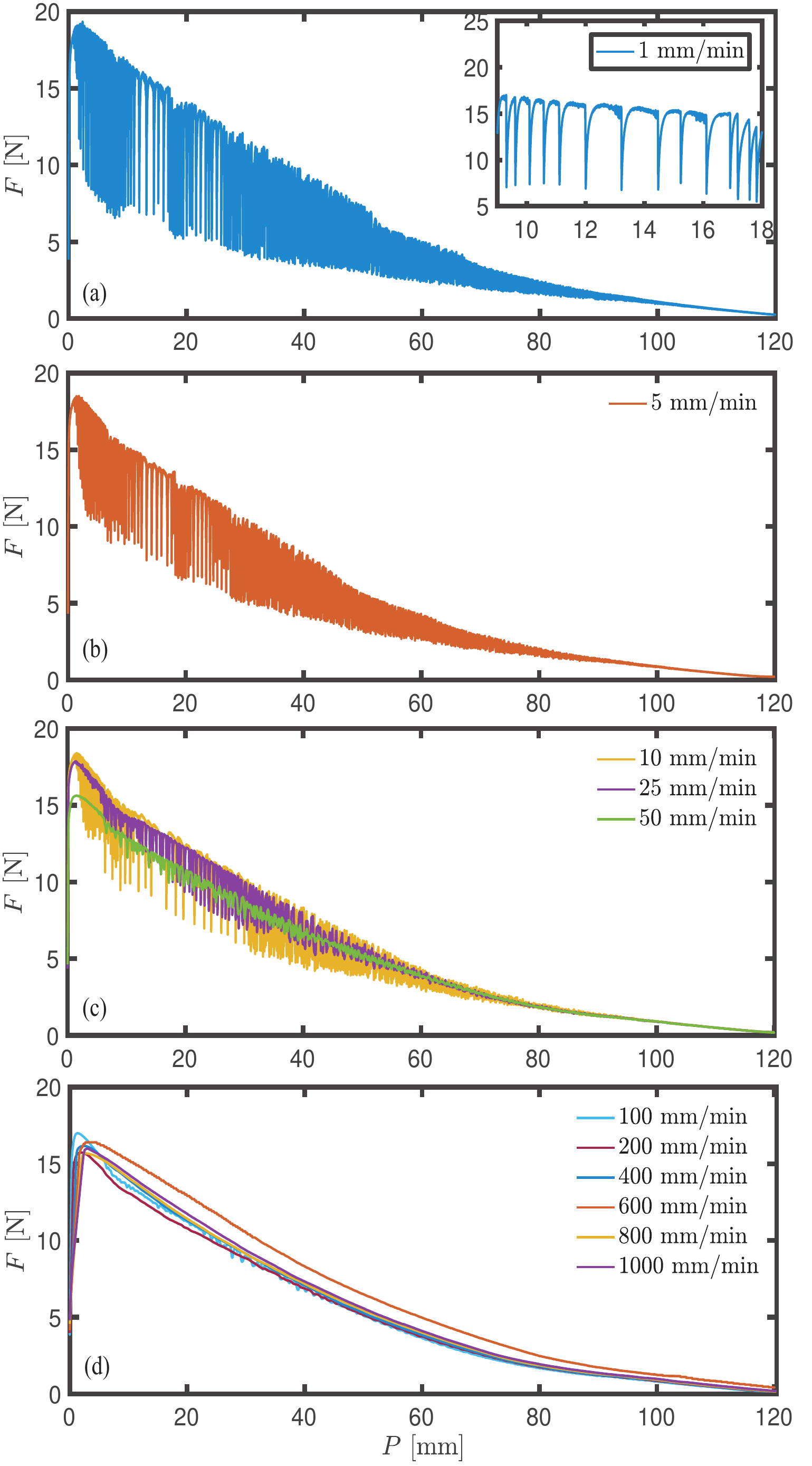}}
\caption{Drag force evolution during uplift tests performed at different uplift velocities (grain size d=\SI{50}{\micro\meter}). $F$ is the measured drag force and $P$ the vertical displacement of the plate relative to its initial position. Velocities are given in the legends. Inset on (a) shows a zoom on $F(P)$ in a small displacement range.}
\label{var_vel}      
\end{figure}

\begin{figure}[h!]
\includegraphics[width=1\columnwidth,trim={0cm 0cm 0cm 0cm}, clip]{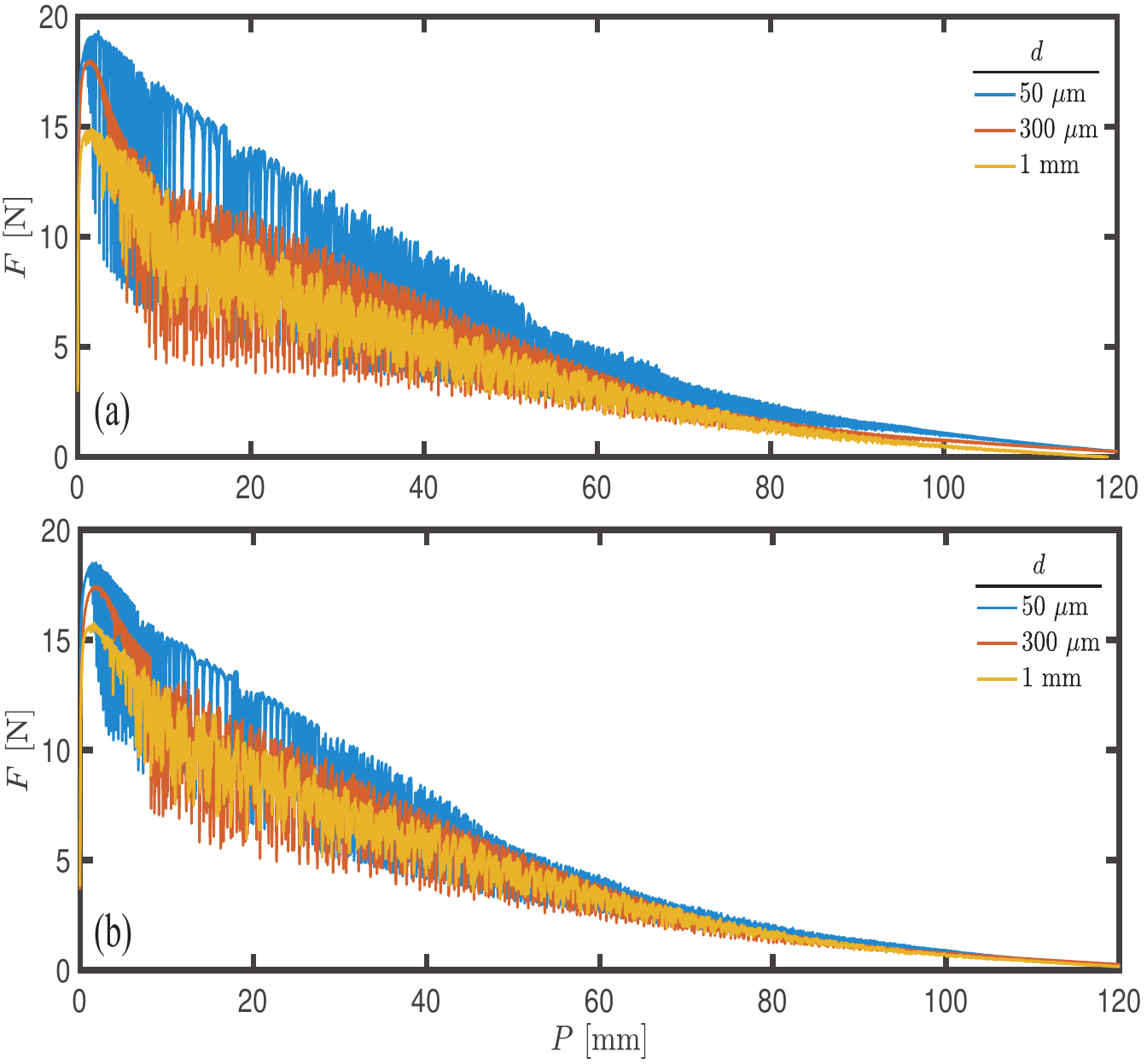}
\caption{Drag force evolution during uplift tests performed with different grain sizes (\SI{50}{\micro\meter}, \SI{300}{\micro\meter} and \SI{1}{mm}) and at two uplift velocities: (a) $v$ = \SI{1}{mm/min} and (b) $v=$ \SI{5}{mm/min}.}
\label{var_grainsizes}       
\end{figure}

\section{Drag force fluctuations}\label{sec:empirical}

We used the experimental method described in the previous Section to measure the drag force evolution in uplift tests performed at different uplift velocities using granular materials of different grain size. Table \ref{tab:parameter} summarises the parameters of these tests. In this section, we seek to characterise the fluctuations in drag forces empirically. The analysis of their physical origin is presented in Section \ref{sec:discusion}.

\subsection{Drag force evolution}

Figure \ref{var_vel} shows the drag force measured in tests performed with the smallest grain size (d= \SI{50}{\micro \meter}) and with different uplift velocities. All force-displacement curves exhibit a similar shape: the force first rapidly increases, reaches a maximum  which we refer to as pullout capacity $F_0$, and then gradually decreases as the plate approaches the surface. It appears that the value of the pullout capacity $F_0$ is only marginally affected by the velocity: it slightly decreases from about \SI{18}{N} to \SI{16}{N} as the uplift velocity is increased. 

\noindent  At slow uplift velocities, large fluctuations in drag forces develop after the peak drag force is reached. The slowest test leads to the largest fluctuations (figure  \ref{var_vel}a), with drag force repeatedly dropping from about $\SI{17}{N}$ down to $\SI{7}{N}$. This corresponds to a drop amplitude of $(17-7)/17\approx 60\%$. At higher uplift velocities, the amplitude of these fluctuations decreases (figures  \ref{var_vel}a-c). There is no visible fluctuation for velocities higher than about \SI{100}{mm/min} (figures  \ref{var_vel}d). This indicates that the mechanism leading to drag instability is strongly rate-dependent.

Figure \ref{var_grainsizes} shows the drag forces measured in tests conducted with different grain sizes and at two different uplift velocities. Results indicate some variations in uplift capacity from about $\SI{17}{N}$ for the smallest grain size to $\SI{15}{N}$ for the largest grain size. We attribute these variations to possible differences in grain size distribution and grain shape between the three granular materials, which in turns would lead to small differences in internal friction angle. Drag fluctuations occurred with all grain sizes. However, the drag fluctuations magnitude is seemingly smaller with larger grains. 

\begin{figure}[!t]
  {\includegraphics[width=1\columnwidth]{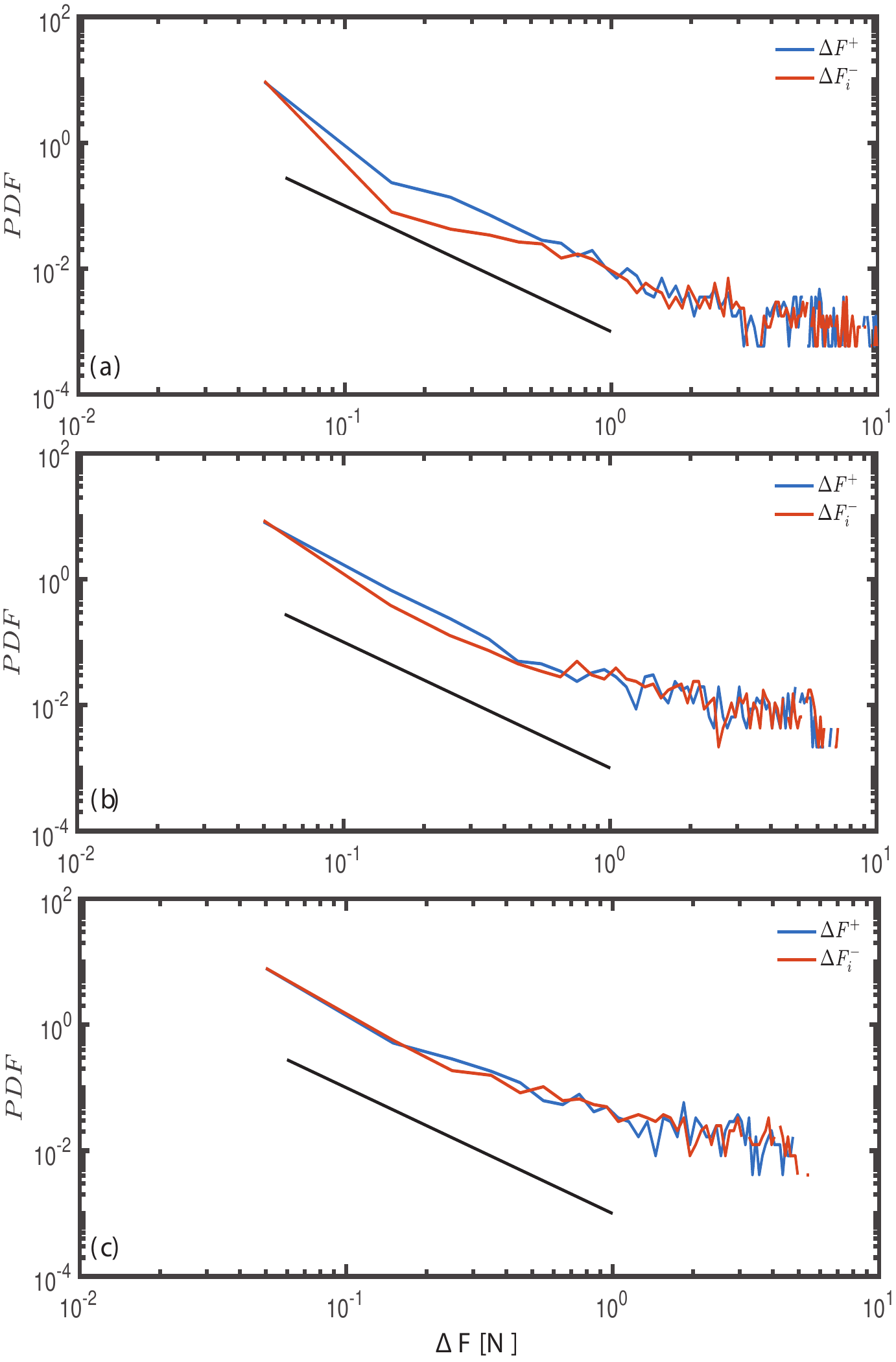}}
\caption{Probability density function (PDF) of the magnitude of drag drops ($\Delta F^-$) and growth ($\Delta F^+$) events, counted throughout uplift tests performed with \SI{50}{\micro\meter} grains at different uplift velocities: (a) \SI{1}{mm/min} (b) \SI{5}{mm/min} and (c) \SI{10}{mm/min} intruder speeds. The black line represents a power law of exponent $-2$ for visual reference.}
\label{pdf50}       
\end{figure} 

\begin{figure}[htb!]
  {\includegraphics[width=1\columnwidth]{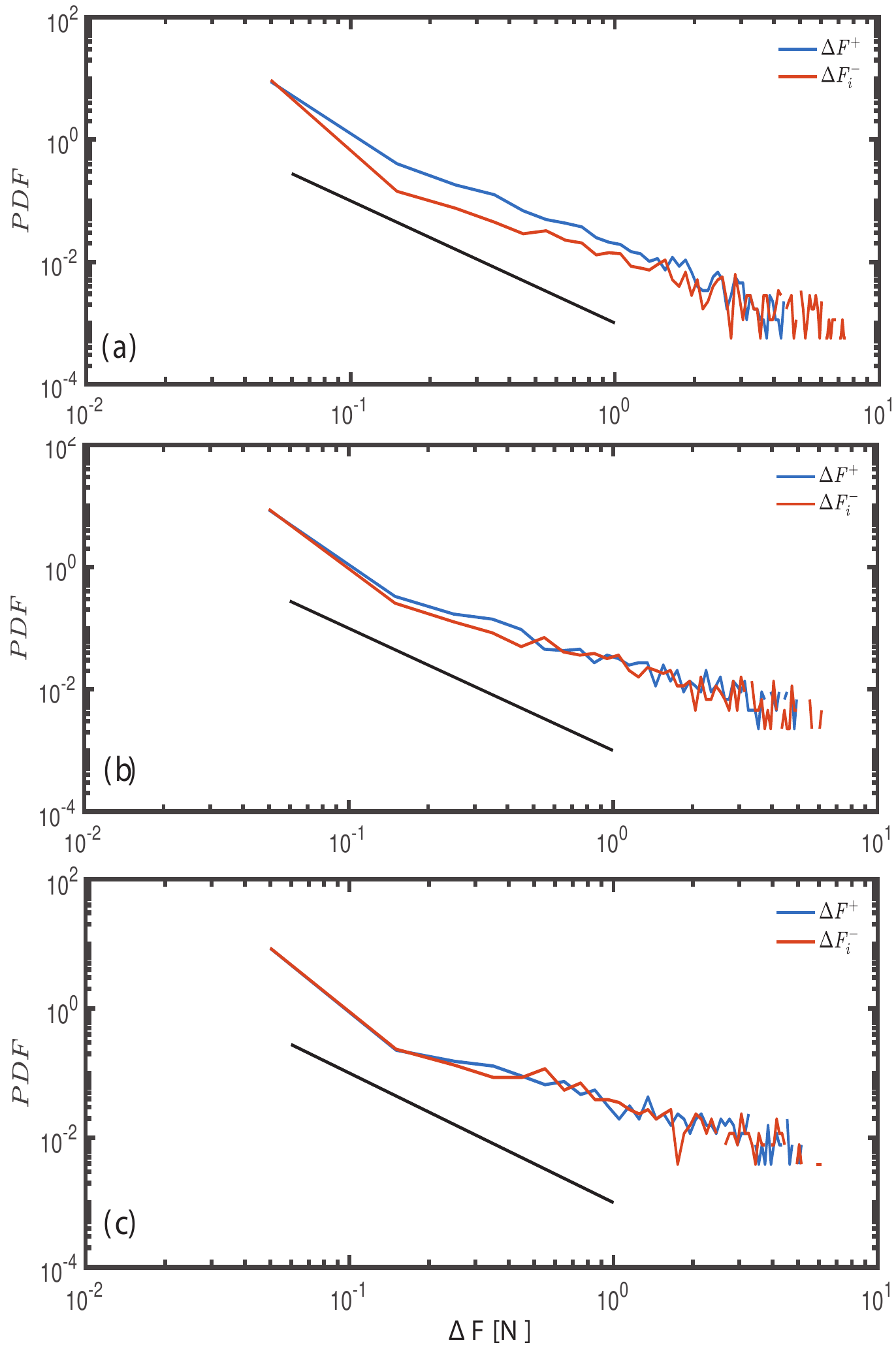}}
\caption{Probability density function (PDF) of the magnitude of drag drops ($\Delta F^-$) and growth ($\Delta F^+$) events, counted throughout uplift tests performed with \SI{300}{\micro\meter} grains at different uplift velocities: (a) \SI{1}{mm/min} (b) \SI{5}{mm/min} and (c) \SI{10}{mm/min} intruder speeds. The black line represents a power law of exponent $-2$ for visual reference.}
\label{pdf300}       
\end{figure}

\subsection{Empirical characterisation of drag fluctuations}

\subsubsection{Distribution in drag drop \& growth magnitude}

We first measured the distribution in drag drop and growth magnitude during each test. Drag drop and growth events are recorded after the uplift capacity is reached, which corresponds to the onset of drag fluctuations. They are identified by finding the series of displacements $P_k$ at which the derivative $\partial F/\partial P$ changes sign - $P_k$ denotes the $k$-iest element of this series. We computed $\partial F/\partial P$ from the discrete time series of $F$ using a forward finite difference scheme. This procedure enabled us to measure the magnitude of the drag force variations between two consecutive extrema:

\be
\Delta F_k = |F(P_{k+1})-F(P_k) |
\ee

\noindent Negative values of  $F(P_{k+1})-F(P_k)$ correspond to a drop in drag force while positive values correspond to a growth in drag force. We refer to these as $\Delta F^-$ and $\Delta F^+$, respectively.
Tests performed at low uplift velocities are typically comprised of a succession of a few thousands of such drop and growth events. 

Figures  \ref{pdf50} and  \ref{pdf300} show the probability density function (PDF) of $\Delta F^{\pm}$ for tests performed at different uplift velocities, with either \SI{50}{\micro\meter} or \SI{300}{\micro\meter} grains.  Results evidence a wide distribution in drag force drop and growth magnitudes, which spans over about two decades from $\SI{0.1}{N}$ to $\SI{10}{N}$. Results also evidence a similarity between the distributions $\Delta F^+$ and $\Delta F^-$. This indicates that drag force drops and grows in a similar manner. This rules out a hypothetical scenario whereby the drag force would recover from a large drop by a succession of smaller drop and growth events.

The PDF of drop and growth magnitude is consistent with a power law of exponent $-2$, $\text{PDF}(|\Delta F|)\propto |\Delta F|^{-2}$. This indicates that there are fewer large events than small events. For large events, data indicate a deviation from this power law and a plateauing. This can be attributed to a poor statistical representation associated with the limited number of such events during any given test. PDFs are further characterised by a maximum value of $|\Delta F|$, above which no event is detected. Figure  \ref{pdf50} and  \ref{pdf300} show that this cutoff value becomes smaller at larger uplift velocity: it is approximately $\SI{10}{N}$ at \SI{1}{mm/min} and \SI{7}{N} at \SI{5}{mm/min}. This indicates that increasing the uplift velocity affects the existence and number of large events, but does not significantly affect smaller events. 

In order to check the effect of the noise inherent to our experimental system, we have performed a similar analysis for a plate that is kept immobile in the granular packing. The result evidenced a constant force with fluctuations of $\pm$ \SI{0.01}{N}. This value is significantly lower than the smallest detected fluctuations on figure \ref{pdf50} and \ref{pdf300}, which are about \SI{0.05}{N}. This suggests that the measured fluctuations are not directly affected by the system noise.

\begin{figure}[tb!]
 {\includegraphics[width=1\columnwidth,]{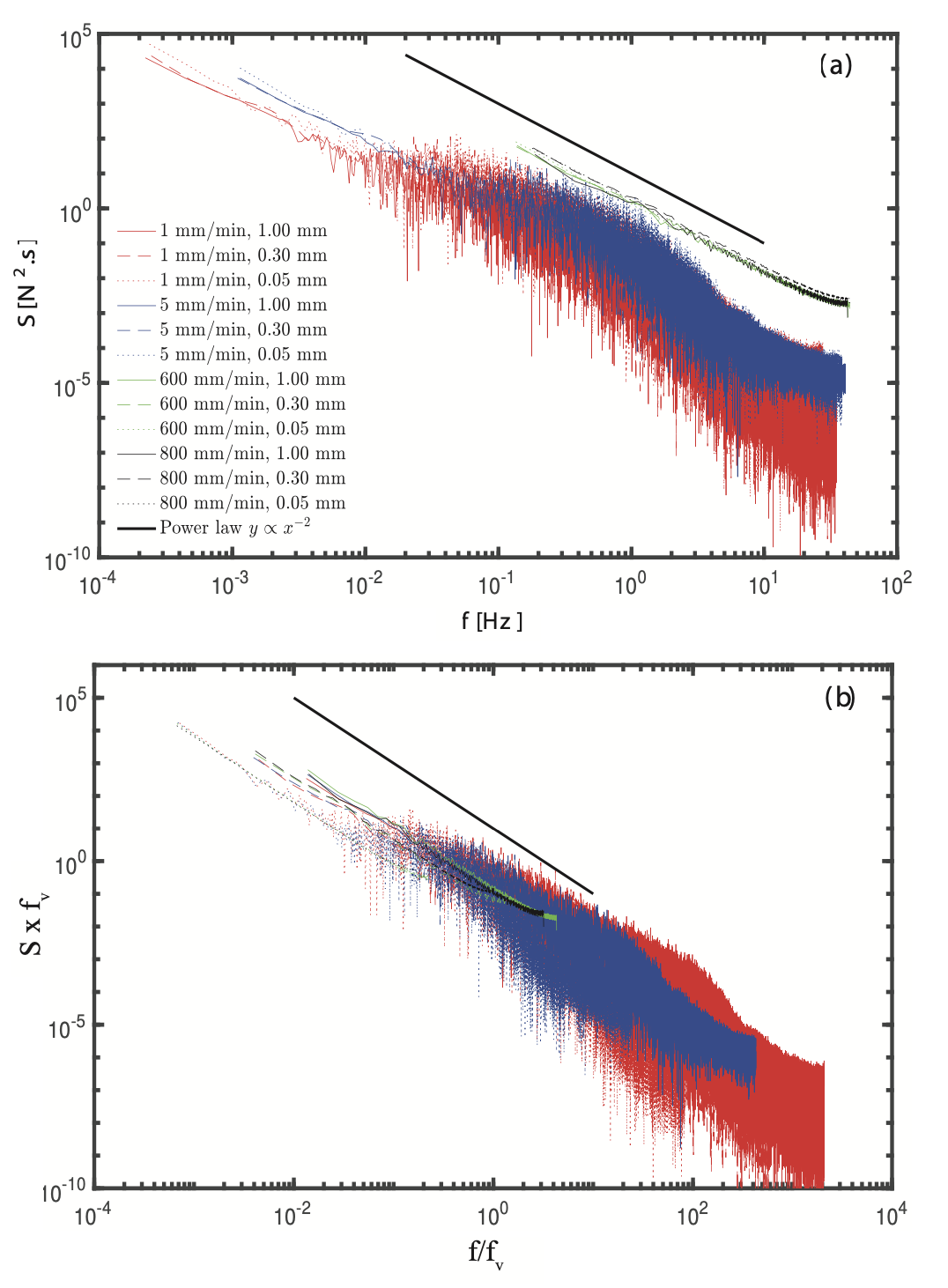}}
\caption{Power spectra $S$ of the drag force time series measured in uplift tests performed with different grain sizes and uplift velocities. (a) Power spectra $S(f)$ measured as per Eq. (\ref{eq:PS}); (b) Same power spectra normalised by $f_v$, a characteristic frequency of the system. Black lines represents a power law with an exponent $-2$ for visual reference; the other lines are experimental data (see legend in (a)).}
\label{power_exp}       
\end{figure}

\subsubsection{Power spectral density}

We characterised the drag fluctuation frequency by measuring the power spectral density $S$ of the force-displacement curves, defined as:

\be \label{eq:PS}
S (f) = \frac{\Delta t^2}{T} \left | \sum_{n=1}^N F_n e^{-i f n \Delta t}\right|^2
\ee

\noindent where $F_n$ is the drag force measured at time $t = n\Delta t$, $\Delta t$ is the period between to consecutive force readings, and $T=N \Delta t$ is the total duration of the test. $S(f)$ was estimated using a fast Fourier Transform algorithm. The physical dimension of $S(f)$ is force squared times time. Accordingly, $\sqrt{S(f)}$ is a measure of the relative magnitude of the drag force fluctuations occurring at a frequency $f$. 

Figure \ref{power_exp} shows the power spectral density measured in tests performed at different uplift velocities and with different grain sizes. The maximum frequency is controlled by the recording rate of the drag force. The minimum frequency is controlled by the total duration of the test. Figure  \ref{power_exp}a shows that $S (f)$ approximately follows a power law $S \approx S_0 f^{-2}$, with $S_0=S(f=1)$ a characteristic power density which depends on the uplift velocity. A similar power law was found when analysing the power spectra of force or stresses different granular systems \cite{demirel1996friction,rozman1996stick,geng2005slow}. 
Figure \ref{power_exp}b shows that same data rescaled by an elementary characteristic frequency $f_v$ of the system, defined as:
 
\be
f_v = \frac{v}{d} = t_v^{-1}
\ee

\noindent $f_v$ corresponds to the frequency at which the plate moves up by a distance of one grain size $d$. The collapse of the rescaled power spectral densities indicates the following scaling:

\be
S(f) = \zeta \frac{1}{f_v} \frac{f^2_v}{f^2} = \beta \frac{f_v}{f^2}
\ee

\noindent with $\zeta$ a constant with a dimension of a force square, which does not depend on the grain size or uplift velocity. Given its physical dimension, this constant could be related to the uplift capacity $F_0$. 

This collapse suggests that an elementary frequency of granular drag fluctuations is $t_v$, which corresponds to how often the plate moves up on a distance of one grain size.  

\begin{figure}[!t]
 {\includegraphics[width=1\columnwidth,trim={0cm 0cm 0cm 0cm}, clip]{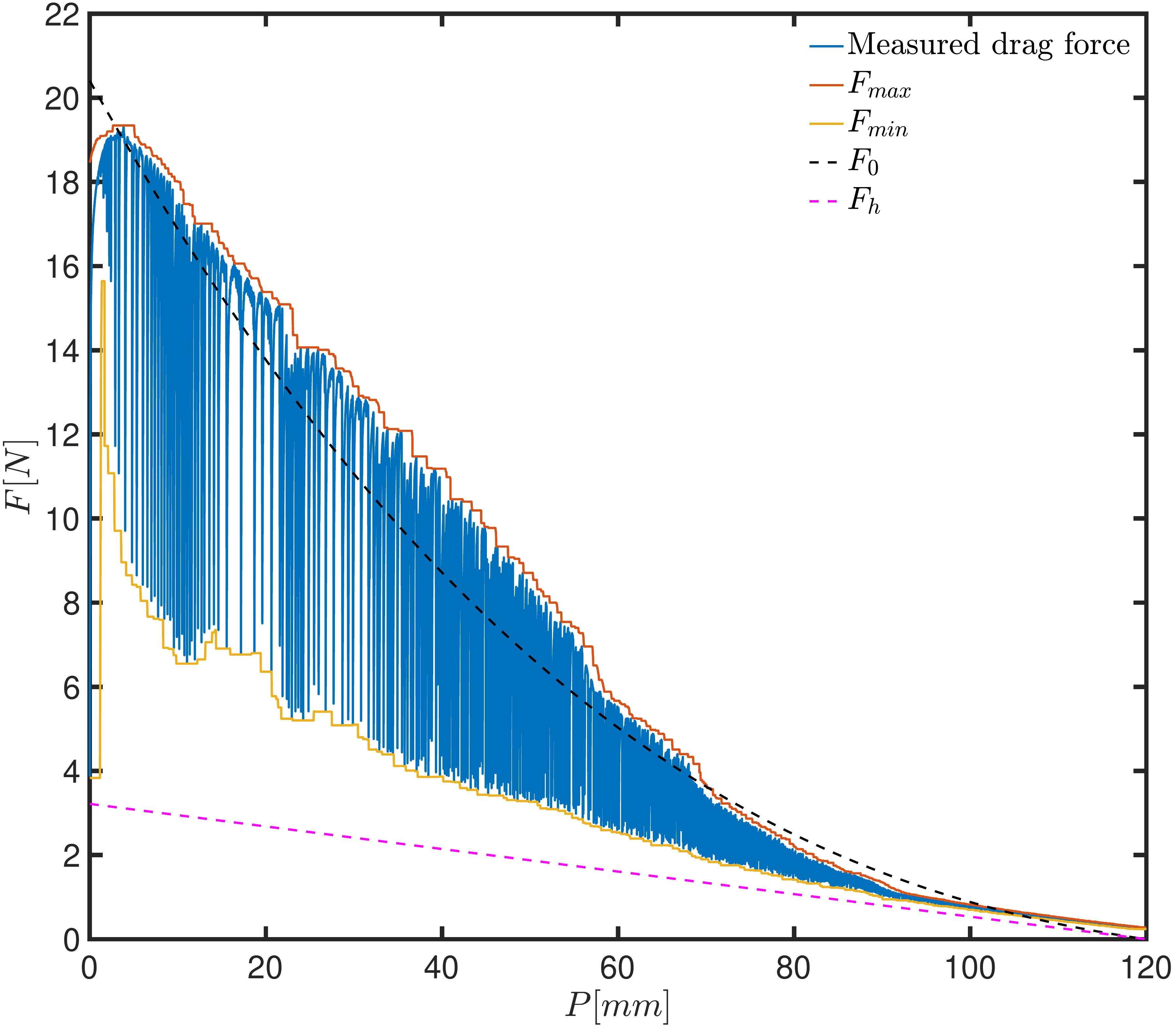}}
\caption{Upper and lower bounds for the amplitude fluctuations. Example of a force-displacement curve measured with $d=$\SI{50}{\micro\meter} at  $v=1$ mm/min. The red and yellow lines corresponds to the local maxima and minima of $F(P)$ (see text). For comparison, dashed lines show the characteristic forces  associated with hydrostatic pressure $F_h(H-P)$ (eq. \ref{eq:Fh}) and with the weight of a truncated cone $F_0 (H-P)$ (eq. \ref{eq:F0}).}
\label{fig:mov_avg}       
\end{figure}


\begin{figure}[h!]
 {\includegraphics[width=1\columnwidth]{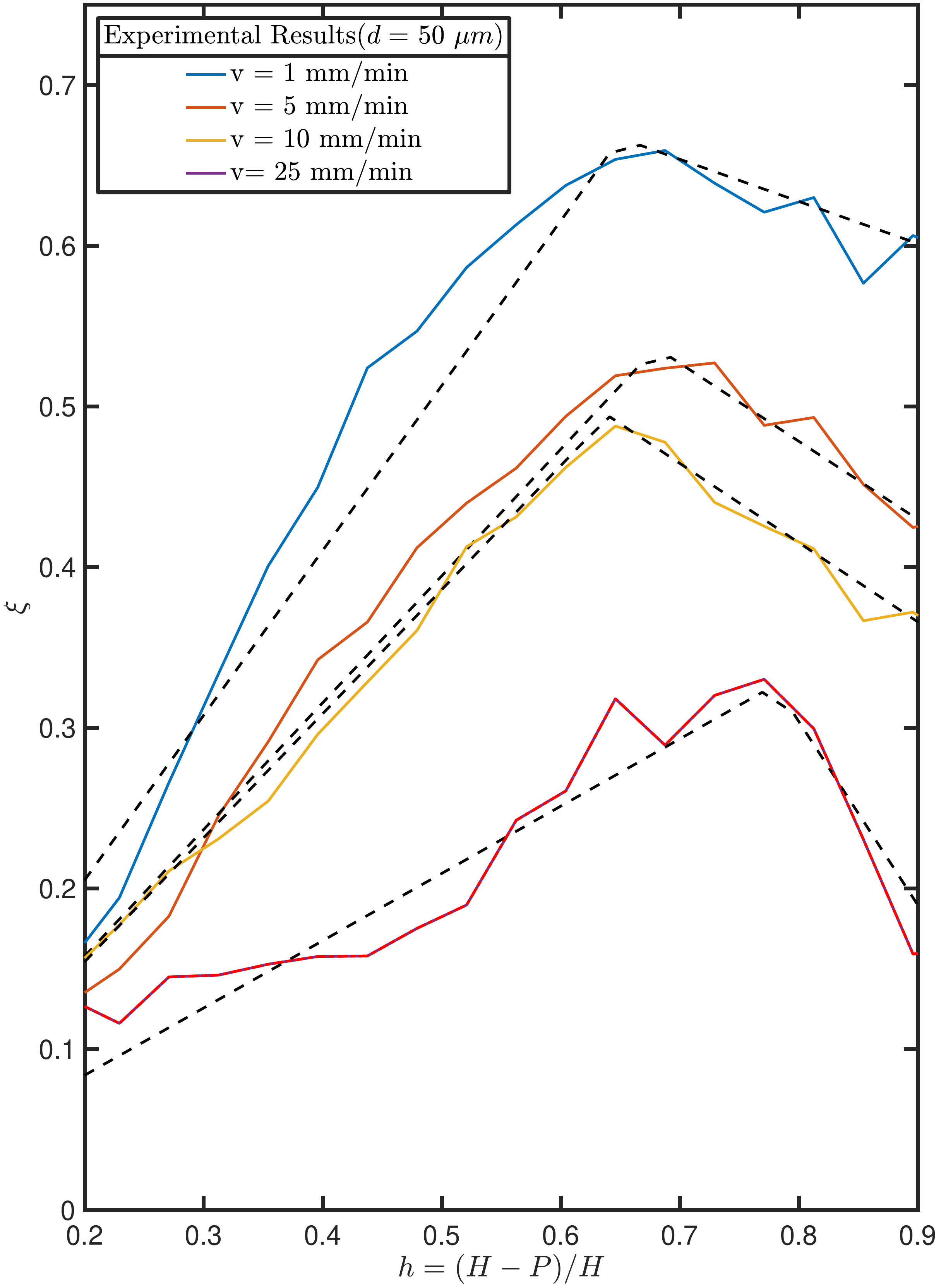}}
\caption{Drag fluctuation amplitude $\xi$ (defined in eq. \ref{eq:xi}) as a function of the plate position $h$ during uplift tests performed with $d=$\SI{50}{\micro\meter} and at different uplift velocities. Dashed lines represent the best fits using the piecewise linear function defined in Eq. (\ref{eq:empirical_Fluc}).}
\label{multiple_ampl}       
\end{figure}

\begin{figure}[h!]
 {\includegraphics[width=1\columnwidth]{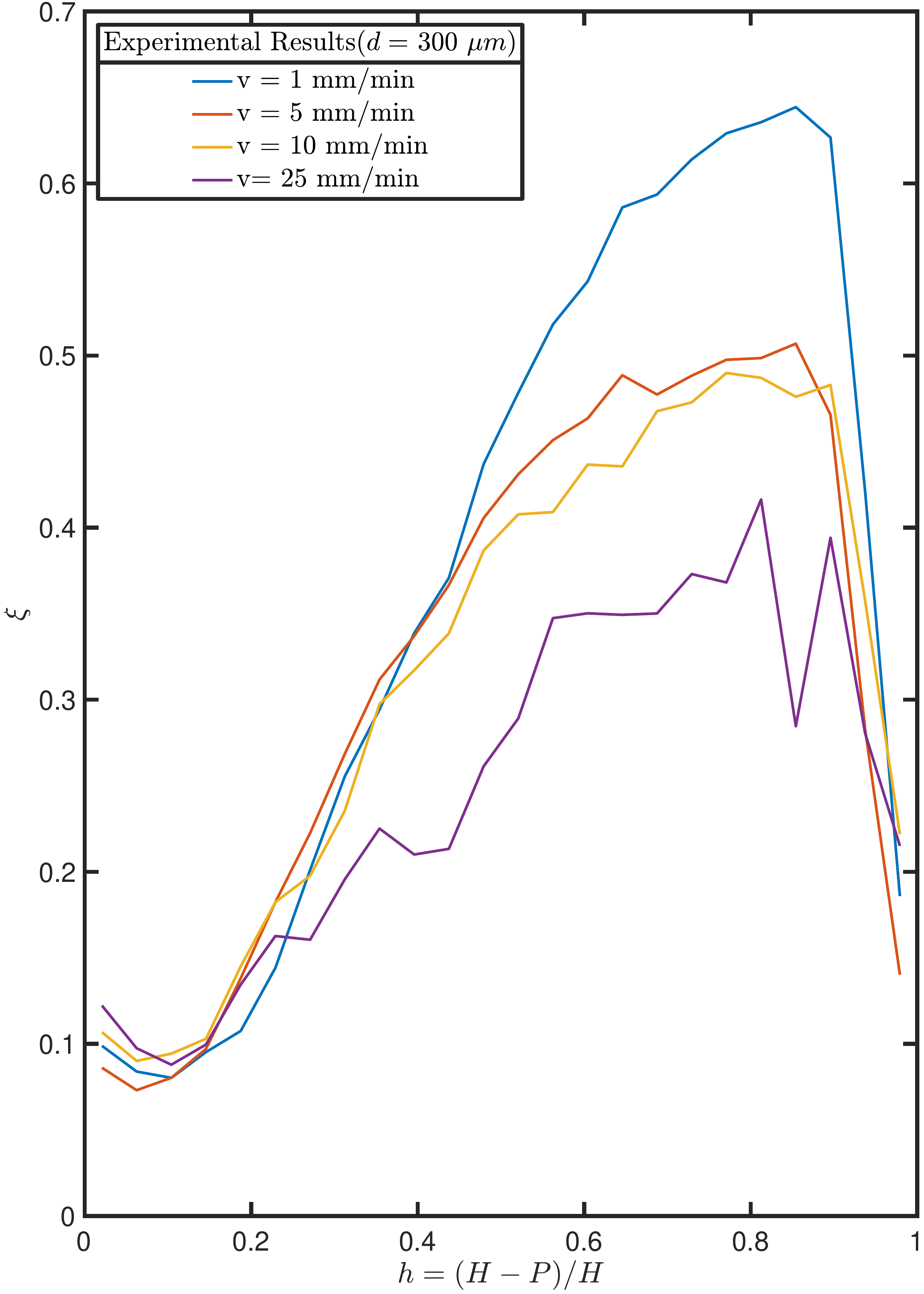}}
\caption{Drag fluctuation amplitude $\xi$ (defined in eq. \ref{eq:xi}) as a function of the plate position $h$ during uplift tests performed with $d=$ \SI{300}{\micro\meter} and at different uplift velocities. Dashed lines represent the best fits using the piecewise linear function defined in Eq. (\ref{eq:empirical_Fluc}).}
\label{multiple_300}       
\end{figure}

\begin{figure}[h!]
 {\includegraphics[width=1\columnwidth,trim={0cm 0cm 0cm 0cm}, clip]{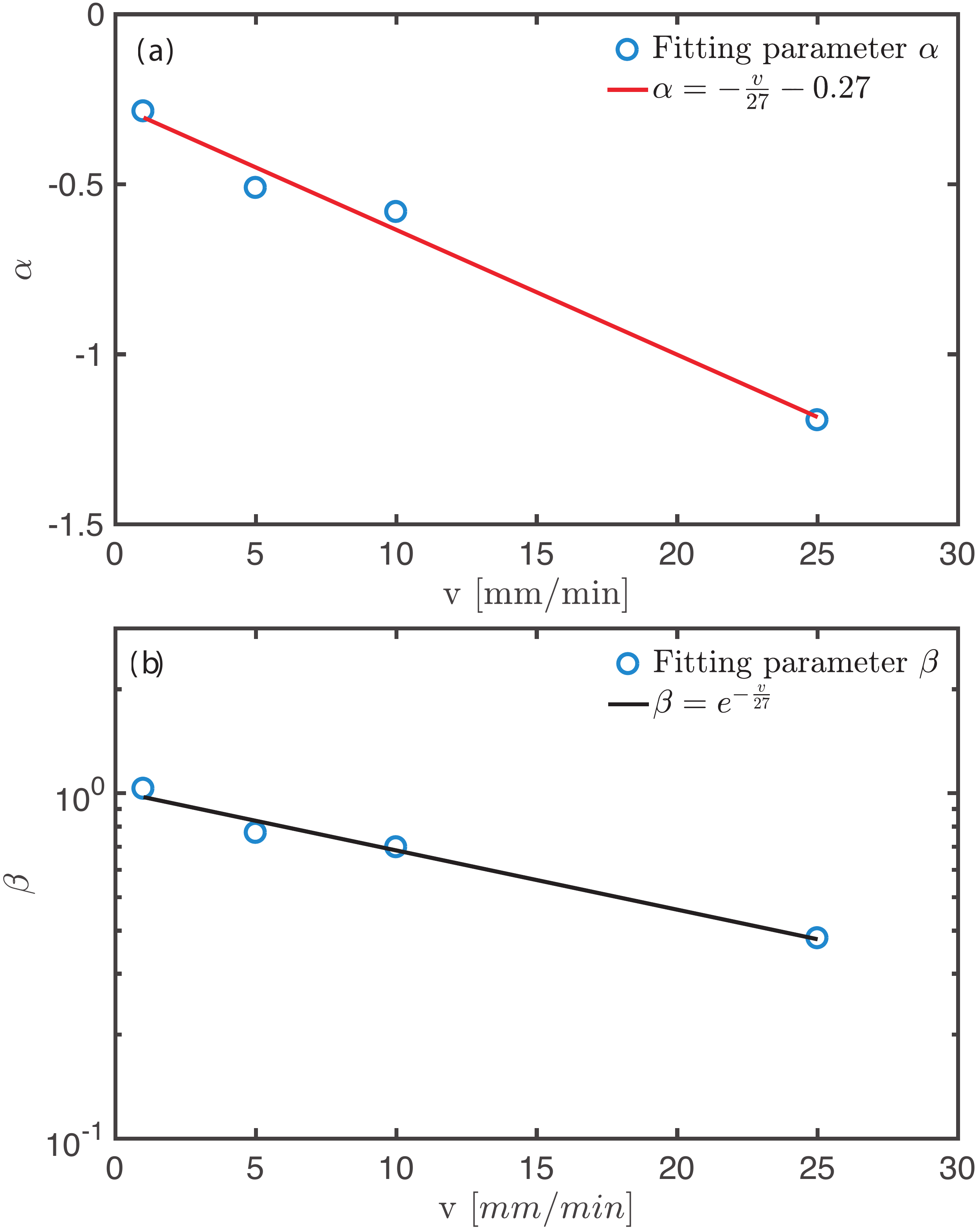}}
\caption{
Value of the fitting parameters for the amplitude fluctuation $\xi(h)$, corresponding to tests with $d=$\SI{50}{\micro\meter}. (a) and (b) shows the two slopes of the piecewise linear function (\ref{eq:empirical_Fluc}), fitted to the experimental data shown on figure \ref{eq:empirical_Fluc}. A linear and exponential functions are shown for comparison.
}
\label{slope_final}       
\end{figure}

\subsubsection{Evolution of fluctuation magnitude during uplift}

The previous methods for characterising drag fluctuations considered the drag force evolution during a full test. However, figures \ref{var_vel} and
\ref{var_grainsizes} suggest that the magnitude of the drag fluctuation varies as the plate gets closer to the surface. As a way to characterise this evolution, we measured the following quantity:

\be \label{eq:xi}
\xi(P) = 1 - \frac{F^{min}(P\pm\delta)}{F^{max}(P\pm\delta))}
\ee

\noindent  where $F^{max}(P\pm\delta)$ and $F^{min}(P\pm\delta)$ are the minimum and maximum drag forces measured within a displacement interval $P\pm \delta$.

Figure \ref{fig:mov_avg} illustrates the result of this procedure for a test performed with the smallest grains and at the slowest velocity. $F^{max}$ and $F^{min}$, measured using $\delta=$\SI{2.5}{mm},  form an upper and lower bound for the drag force $F(P)$. They capture both the large magnitude fluctuations and the slow decrease in drag force as the plate gets closer to the surface.

Figure \ref{multiple_ampl} shows the evolution of the drag force fluctuation magnitude measured by $\xi$ in tests performed at different uplift velocities using \SI{50}{\mu m} grains. Results are shown as  a function of the normalised plate distance to the free surface, defined as: 

\be
h = \frac{H-P}{H}
\ee
\noindent Accordingly, $h =1$  corresponds to the initial position of the plate and $h=0$ corresponds to the plate reaching the free surface. Results are shown in the range $0.2 \leqslant  h \leqslant 1$, excluding shallow conditions ($h<0.2$ ) where the plate displacement induces significant surface deformations. 

All tests yield a similar shape for the drag fluctuation $\xi(h)$. The fluctuation magnitude first increases just after the uplift capacity is reached. It then reaches a maximum and gradually decreases as the plate approaches the free surface. We propose to use a piecewise linear function to capture this behaviour, which involves three parameters:

\be \label{eq:empirical_Fluc}
\xi(h) = 
\begin{cases}
    \beta h ,& \text{if } h<h_{max}\\
    \beta h_{max} + \alpha (h-h_{max})  & \text{otherwise.}\\
\end{cases}
\ee

\noindent $\alpha (<0)$ and $\beta (>0)$ are the slopes before and after the maximum fluctuation is reached, and $h_{max}$ is the displacement at which the maximum fluctuation amplitude is reached. Higher order polynomial function could also be used to fit these experimental data.

We have estimated the slopes $\alpha$ and $\beta$ for all tests by fitting the amplitudes of the measured fluctuations $\xi(h)$ by the piecewise linear function \ref{eq:empirical_Fluc}. In this fitting procedure,  $\alpha$ and $\beta$ are free fitting parameters while the value of $h_{max}$ was pre-determined from the measurements by finding the maximum value of $\xi$. Figure \ref{multiple_ampl} shows that this piecewise linear fit captures the measured evolution of the fluctuation amplitude.  Figure \ref{multiple_300} shows that test performed with a larger grain size yielded a similar fluctuation amplitude evolution.

Figure \ref{slope_final} shows that the values of the slopes  $\alpha$ and $\beta$ that best fit the data are strongly dependent on the uplift velocity. $\alpha$'s absolute value increases, approximately linearly, with the uplift velocity $v$ while $\beta$ exponentially decreases with $v$.


 



\section{Inferring the physical origin of drag fluctuations} \label{sec:discusion}

In this section, we discuss the possible origins of the drag fluctuations in light of the experimental observations presented in Section \ref{sec:empirical}. We first assess how drag fluctuations differ from a stick-slip dynamics, and then introduce a physical process compatible with their observed properties.

\subsection{Similarities and differences with a stick-slip dynamics}  

A stick-slip dynamics often develops in granular materials subjected to external loadings \cite{howell1999stress,dalton2001self,otsuki2014avalanche,zadeh2017avalanches}. 
The most basic mechanical analogue that yields a stick-slip dynamics is that presented in figure \ref{fig:stick-slip}a. It involves pushing a linear spring at a constant velocity on 
one end, which is connected on the other end to a mass $M$ and a slider. 

A stick-slip dynamics has been experimentally evidenced while dragging an object in granular materials \cite{albert1999slow,albert2001stick}. In these tests, the loading was applied via an external spring connecting the load cell to the object. The resulting elastic deformation was therefore not directly related to the granular packing mechanical behaviour. In contrast, granular shear experiments evidenced stick slip dynamics without external spring \cite{lherminier2019continuously}. The elastic behaviour was found to directly arise from the compression and buckling of force chains in the packing. This resulting dynamic was found to be closely matching that of earthquakes, including a wide range of slip magnitude.

In our uplift test, one could therefore expect some stick stick-slip to develop. As there is no external spring between the load cell and the plate in our configuration, the elastic element in the analogue would represent packing elastic response. The mass would represent a mass of grains involved in the motion, and the slider would represent the plastic behaviour of the packing. 

To assess whether a simple stick-slip dynamic can capture our observed drag force fluctuation, let us detailed the behaviour of the simple analogue in figure \ref{fig:stick-slip}a. Two conditions are necessary to obtain a stick-slip dynamics with this analogue:

\begin{figure}[tb!]
 {\includegraphics[width=1.1\columnwidth]{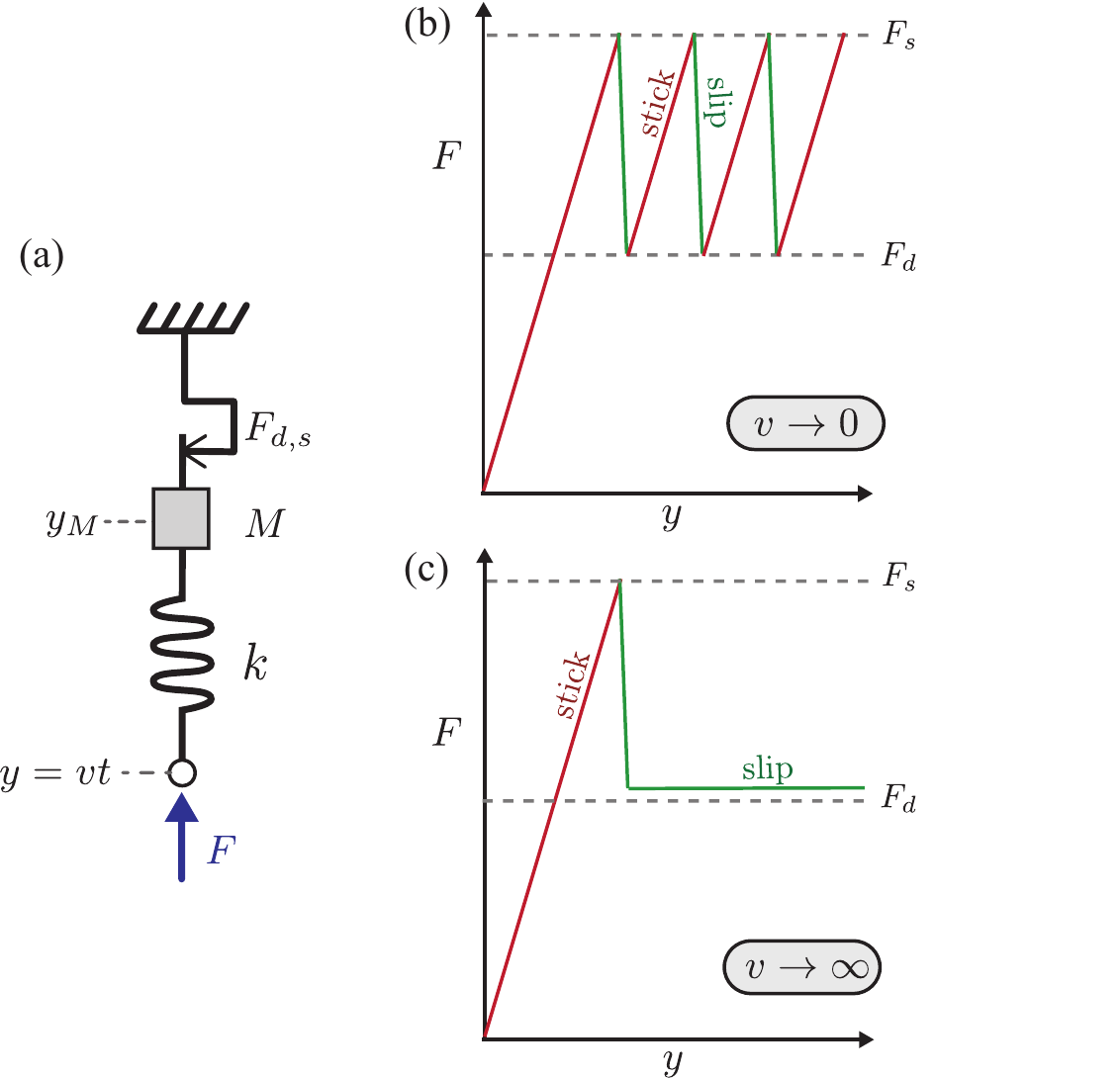}}
\caption{Stick-slip dynamics. (a) Mechanical analogue exhibiting a stick-slip dynamics: a spring of stiffness $k$, a mass $M$ and a slider are connected in series. The point $y$ is driven at constant speed $v$ by applying an external force $F$, while the top-end is fixed. The force in the spring is given by Hooke's law: $\dot F^{spring} = k (\dot y_M-\dot y(t))$; the force in the slider $F^{slider}$ is either $F_d$ when it slides or $F^{spring}$ when it sticks; the mass moves according to an inertial dynamics $M \ddot {y_M} = F^{spring}-F^{slider}$ when the slider slides, and is fixed otherwise. (b,c) illustrations of a force-displacement response of such mechanical analogue subjected to a low or high loading velocity $v$; the stick-slip dynamics develops at low velocity.}
\label{fig:stick-slip}       
\end{figure}

\begin{enumerate}
\item The slider's sliding criteria must involve a dynamic and a static critical force $F_d$ and $F_s$ that are different, with the static force being strictly larger than the dynamic force: $F_s>F_d$. The static force is the maximum force that a slider initially at rest (sticking) can sustain before sliding. The dynamic force is the minimum force a sliding slider can sustain without sticking;
\item The loading velocity must be slow enough.
\end{enumerate}

Figures \ref{fig:stick-slip}b,c illustrate the force-displacement response of this analogue subjected to slow or rapid loadings.

At infinitely slow loading velocities, the force increases linearly in time during the sticking phase as the spring is being compressed on one end, while the other end is fixed by the sticking slider: $\dot F = kv$. Once the static force is reached, the slider slips letting the mass move, which relaxes the spring with an elasto-inertial dynamics similar to a harmonic oscillator. This leads to a relaxation of the force on a typical time $\sqrt{M/k}$ from $F_s$ to $F_d$. Once $F_d$ is reached, the slider sticks and  a new cycle starts. This stick-slips dynamics leads to regular saw-tooth fluctuations in force, which are reminiscent of our measured drag force fluctuations. 

At high loading velocities, the spring is being compressed significantly during the slip phase, such that the spring force cannot relax and reach $F_d$. As a result, the slider can never stick again once it has started slipping. There is therefore no force fluctuation, and the force converges to $F_d$ during the slip. This rate-dependence of the stick-slip force fluctuations is seemingly consistent with the absence of drag fluctuation we observed at high uplift velocities.  

\begin{figure}[h!]
 {\includegraphics[width=1\columnwidth]{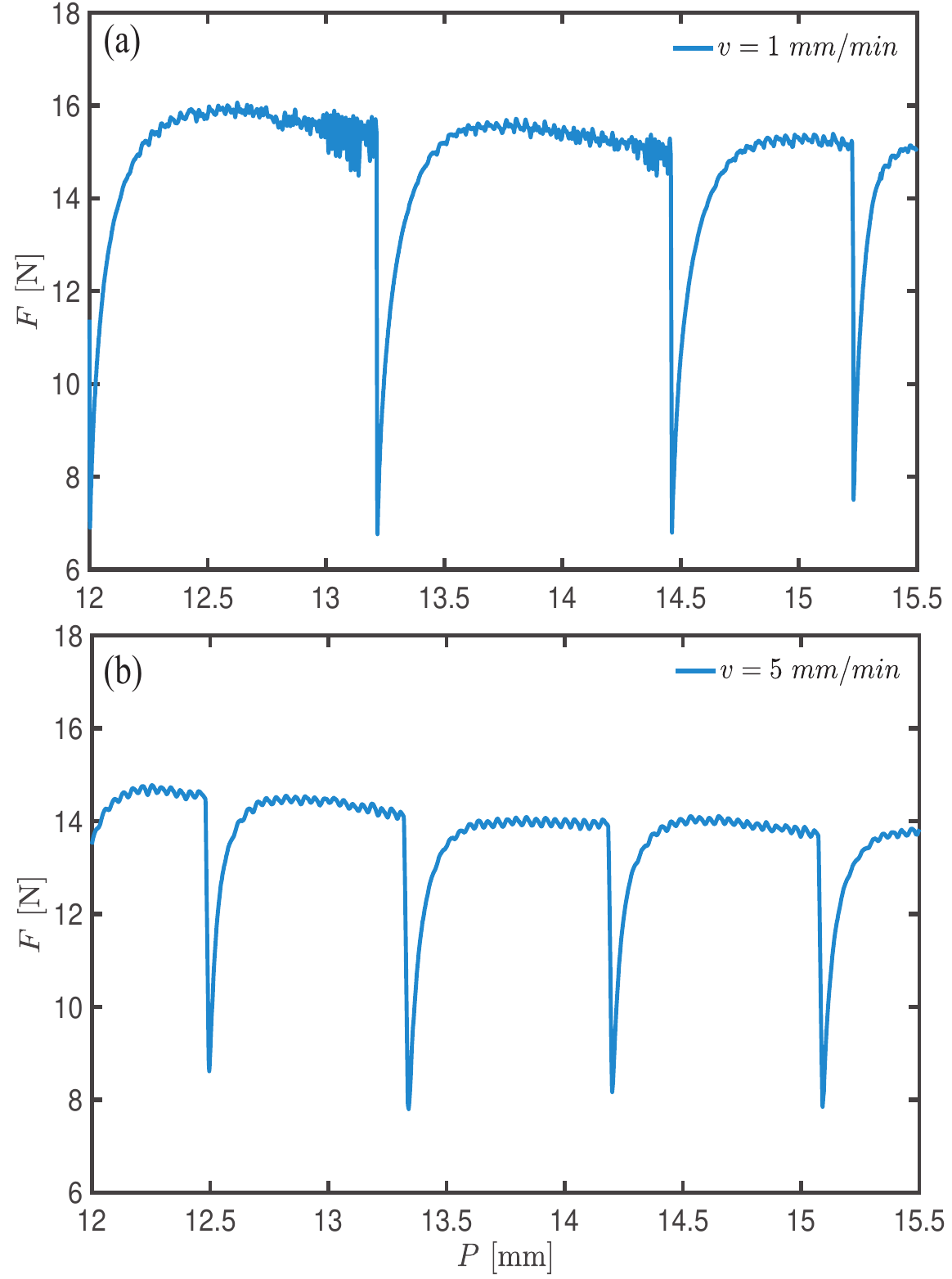}}
\caption{Drag force fluctuations measured with  \SI{50}{\micro\meter} grains at two different uplift velocities, zooming on a small displacement window ranging form about $12$ to $15.5$ mm. These evidence the non-linearity of the force build-up after large relaxation events, the relative regularity of large relaxation events, and the slow relaxation of the drag force between these events.
}
\label{fig:super_zoom}       
\end{figure}

This simple stick-slip analogy, however, does not capture the following features of the granular drag fluctuations:

\begin{itemize}
\item When the drag force increases, it does so in a non-linear way (see figure \ref{fig:super_zoom}). This is not compatible with the linear-spring model and rather suggests some non-linear behaviour.
\item Drag forces then reach a maximum and slowly relaxes, with small fluctuations, before a large relaxation event occurs (see figure \ref{fig:super_zoom}). This challenges the stick-slip analogue assumption that force relaxation (slip) is triggered when a force threshold is reached, as represented by the static force $F_s$. It rather suggests that the criteria for force relaxation may in fact be a threshold in plate displacement. 
\item When there are fluctuations in drag force, the value of the maximum force is depth dependent: it decreases as the plate approaches the surface  (see figure \ref{fig:mov_avg}). This challenges the stick-slip model assumption that $F_s$ is constant. 
\item  At low velocities, the magnitude of the drag force fluctuations may reach about $60\%$. To achieve this amplitude with a slip-stick model, the dynamic force $F_d$ should be about $40\%$ of the static force $F_s$. In this case, the stick-slip model would then predict that, at high velocities, the drag force is constant and equal to $F_d$. In contrast, our results show that the drag force at high velocities would rather be close to the static force $F_s$ (see figure \ref{var_vel}). 
\end{itemize}
 
These observations indicate that a simple stick-slip analogue is not sufficient to explain the drag fluctuation dynamics, even qualitatively. This suggests that this drag instability results from a different physical process.

\begin{figure}[h!]
 {\includegraphics[width=1\columnwidth]{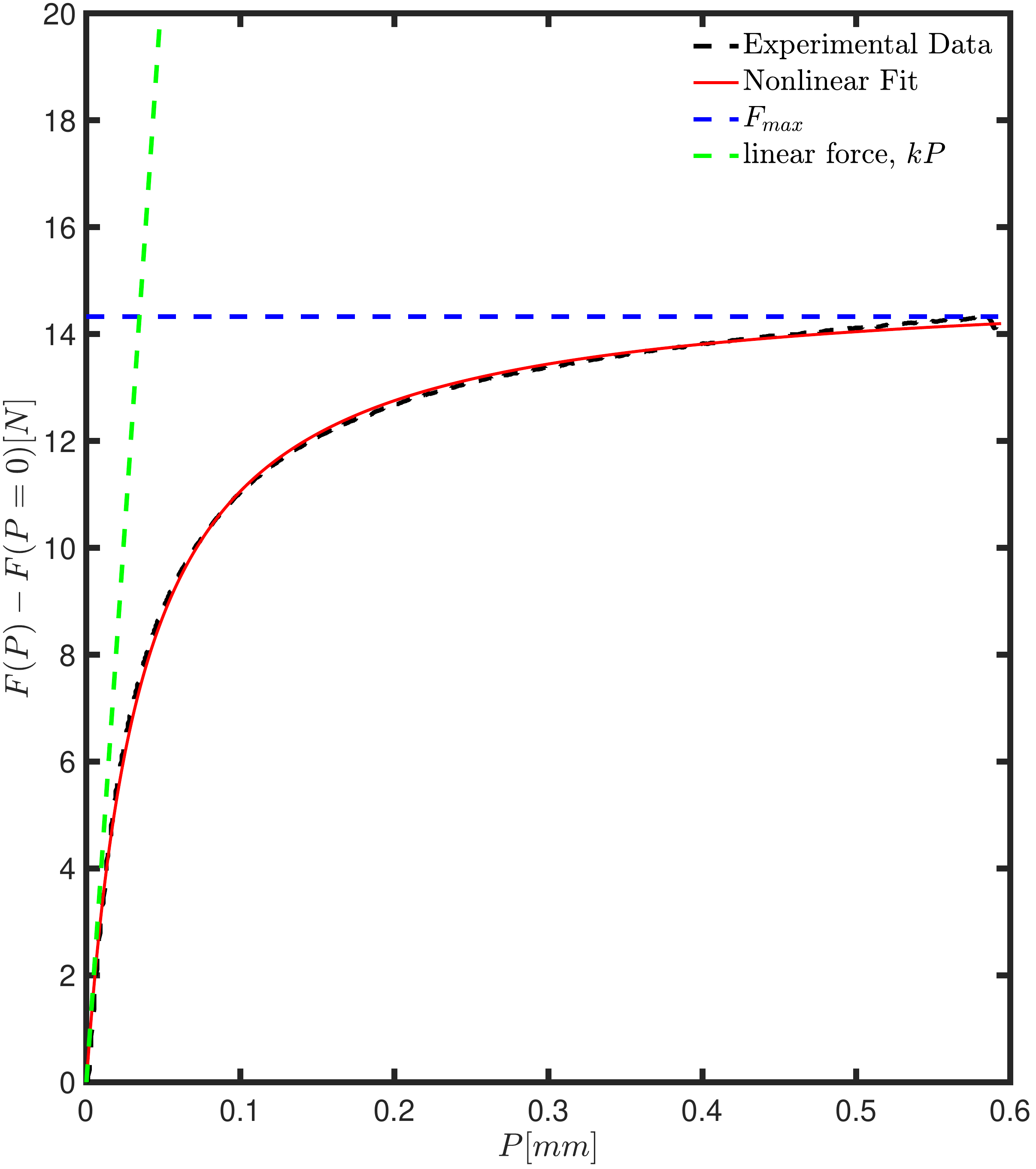}}
\caption{Non linear increase in drag force at the begining of an uplift test: experimental data (d=\SI{50}{\micro\meter} and v = \SI{1}{mm/min}) fitted with equation (\ref{eq:nl}); best fit obtained with $k=4\times 10^4$ N/m and $F_{max}+F(P=0)=1.05 F_0$. (see text).}
\label{nonlin_fit}       
\end{figure}

\subsection{Physical processes controlling granular drag instability}

We present here a number of physical processes that are consistent with the observed feature of the drag instability.

\subsubsection{Non-linearity of the increase in $F(P)$}

The non-linear increase in drag force  $F(P)$ is visible at the beginning of the test, when the drag force first builds up toward $F_0$ (figure \ref{nonlin_fit}). It is also apparent after drag drops, when the drag force increases again (figure \ref{fig:super_zoom}). 
At the beginning of the test, the drag force seemingly linearly increases at small values of $P$, $F(P)\approx F(P=0)+k P$, and then exhibits a non-linear increase toward a maximum $F_{max}$. We propose the following model to capture this behaviour:

\be \label{eq:nl}
F(P) =  F(P=0) + F_{max} \frac{1}{1+\frac{F_{max}}{k P}} 
\ee

\noindent Figure \ref{nonlin_fit} shows that this model captures the measured non-linear increase in drag force, with values of  spring constant $k =4\times10^4$~N/m and a value of $F_{max}$ close to the maximum drag force: $F_{max} \approx F_0-F(P=0)$. This non-linear increase suggests that some plastic deformation may take place before the maximum drag $F_0$ is reached. However, there is no physical rationale for the chosen functional form in (\ref{eq:nl}). Other functional forms such as an exponential decay could equally well capture the experimental data. This initial elasto-plastic response appears to be similar to that of the build up in drag force occurring after large drops (see figure \ref{fig:super_zoom}). This suggests that this build-up results from a similar elasto-plastic process.

A stiffness test is presented in the Appendix, which quantifies the effective stiffness $k_0$ of the system load cell-shaft and plate. Measurements indicate an effective stiffness of   
$k_0 \approx 2.27 \times10^5$ \SI{}{N/m}, which is five times stiffer than the effective stiffness $k$ of the granular packing. This suggests that the measured elastic-like deformation is primarily driven by the granular packing response, with only a marginal contribution of the loading system deformation.

\subsubsection{Maximum drag force}

The maximum drag force $F_0 = \text{max}(F(P))$ is commonly associated to an effective weight of grains located in a volume above the plate. The hydrostatic force $F_h$ arising from hydrostatic pressure considers that this volume is a cylinder above the plate:

\be \label{eq:Fh}
F_h(H) = \phi \rho_g H S.
\ee

\noindent $S = \pi B^2/4 \approx 1.2 \times 10^{-3} \text{ m}^2$  is the surface area of the plate.  
At a depth H=\SI{120}{mm}, this hydrostatic force is $F_h \approx $\SI{3.2}{N}. This is significantly lower than the  measured $F_0$, which are approximately $\SI{19}{N}$ (figures \ref{var_grainsizes} and \ref{var_vel}).

\begin{figure*}[!h]
 {\includegraphics[width=2\columnwidth]{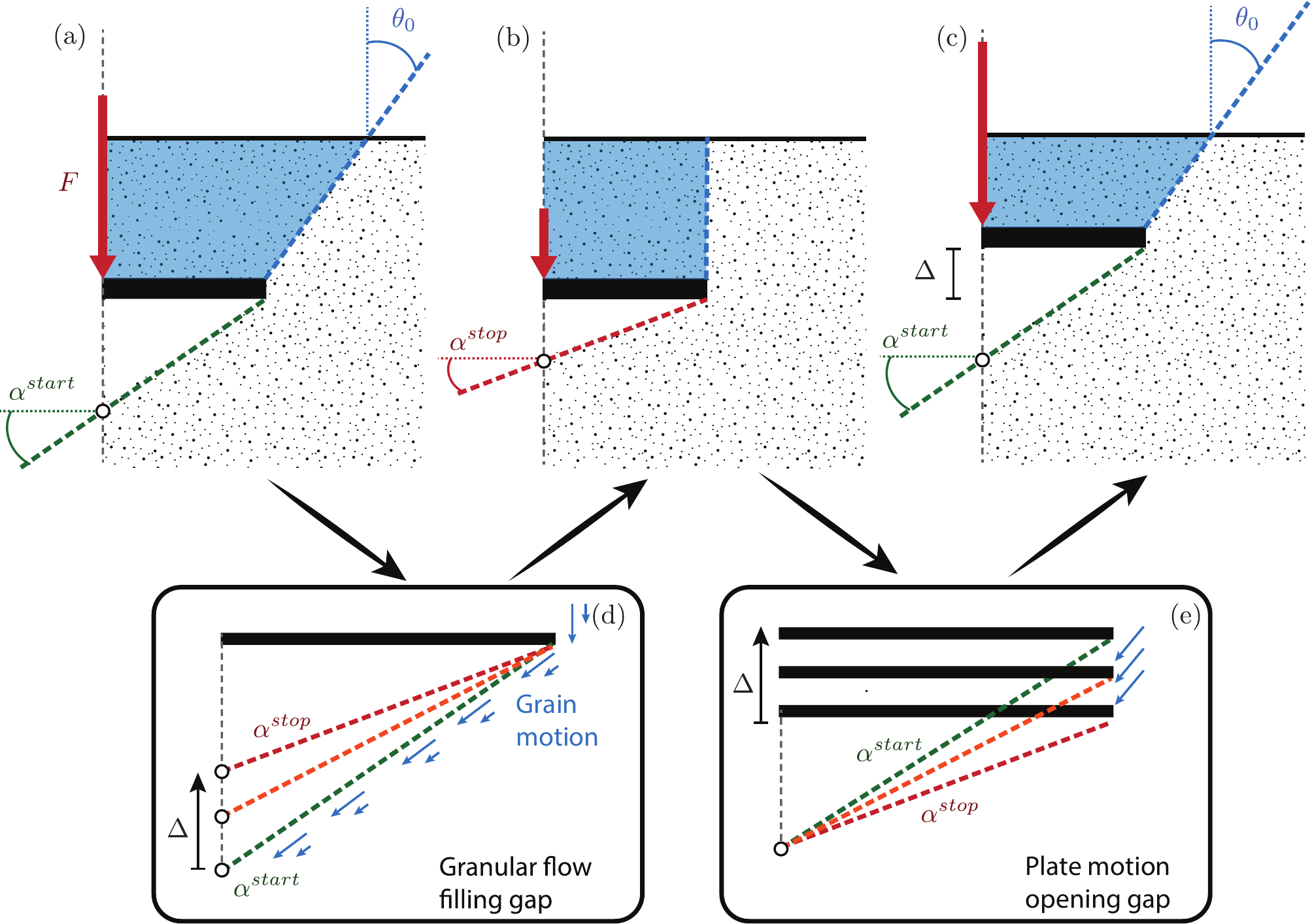}}
\caption{Proposed mechanisms underlying drag fluctuations. (a,b,c) Illustration showing half a plate (black), the grains it supports (blue) and the gap underneath it (white) during a drag force fluctuation cycle. The drag force is maximum in (a) and (c) and minimum in (b). (d,e) Illustration of the proposed \textit{flapping dynamics} underneath the plate, involving a gap opening and a gap filling mechanism}; dashed lines represents the evolution in slope angle of the granular packing free surface; blue arrows represents the direction of the granular flow.
\label{fig:mechanism}       
\end{figure*}

The established explanation is that the volume of grains supported by the plate when the maximum drag is reached is larger than this cylinder. While there are different expressions for the precise shape of this volume, a first approximation is that of a truncated cone, as illustrated in figure \ref{fig:mechanism}a \cite{das2013earth}. Accordingly, the maximum force is given by the following expression:

\bee \label{eq:F0}
F_0(H) &=& F_h f(H); \\
f(H)&=&\frac{1}{3} \left[\left(1+2\frac{H}{B} \tan \theta_0 \right)^2 +2 \frac{H}{B} \tan \theta_0 + 2 \right].
\eee

\noindent The function $f$ represents the ratio of volumes of (i) the truncated cone and (ii) the cylinder of diameter $B$ and height $H$. It is therefore always greater than $1$.  It involves the cone angle $\theta_0$. This angle is usually related to the friction angle of the granular packing \cite{das2013earth,meyerhof1968ultimate,rowe_behaviour_1982,murray_uplift_1987,merifield2006ultimate,kumar2008vertical},  object shape \cite{dyson2014pull,khatri2011effect,bhattacharya2014pullout,askari2016intrusion,giampa2018effect} and the grain size \cite{sakai1998particle,athani2017grain,costantino_starting_2008}. However, there is no general model available to predict it. In order to match $F_0=$\SI{19}{N}, $\theta_0$ would need to be close the internal friction angle of our glass beads packing: $\theta_0 \approx \phi=23^\circ$.

Beyond a critical embedment ratio $H/B$, the cone model is no longer valid as the failure planes do not reach the free surface; the function $f(H)$ then plateaus. Typical values of critical embedment ratios are of the order $5$ but can vary with the type of material \cite{das2013earth}. 

\subsubsection{A flapping dynamic above the plate}

The model of maximum drag force $F_0$ presented above was semi-empirically derived to capture the first maximum force one can apply to the plate before it moves through the packing. It is therefore a failure criterion, and was not intended to capture the drag force behaviour after the peak force is reached. Nonetheless, we propose here to use this failure model to help explain the behaviour of the drag force throughout the uplift. Specifically, it helps defining two elementary forces $F_h(H-P)$ and $F_0(H-P)$ during the uplift, with $F_0>F_h$. This means that the drag force could possibly fluctuate between  these lower and upper bounds. Accordingly, the maximum magnitude of the drag force fluctuations, measured by the ratio of these two forces, would be: 

\be
\frac{F_0(H-P)}{F_h(H-P)}=f(H-P). 
\ee

\noindent This assumption is consistent with our observations in two respects. Firstly, this ratio decreases and tends to $1$ as the plate gets closer to the surface. This is consistent with the decrease in fluctuations magnitude observed in figures \ref{multiple_ampl} and \ref{multiple_300}. Secondly, figure \ref{fig:mov_avg} shows that the two forces $F_0(H-P)$ and $F_h(H-P)$ approximately bound the measured drag forces throughout the test which yielded the largest fluctuations. We note that $F_h(H-P)$ is significantly lower than the lowest drop drag force, indicating that the drag force does not fully relax toward $F_h$. We also note that the maximum of the drag force marginally exceeds $F_0(H-P)$. This could be due to an increase in internal friction angle of the packing above the plate induced by compaction.

Accordingly, we propose that drag fluctuation occurs when the system switches between two extreme configurations, as illustrated on figures \ref{fig:mechanism}a-c. Before a large drop in drag force, a cone of grains is supported by the plate, leading to a force close to $F_0(H-P)$. After the drop, at least a cylinder of grains is supported by the plate, leading to a minimum force of $F_h(H-P)$. At high velocity, as the fact that the drag force remains close to $F_0(H-P)$ suggests that there is a cone of grain supported by the plate at any point of time.

\begin{figure}[tp!]
 {\includegraphics[width=\columnwidth]{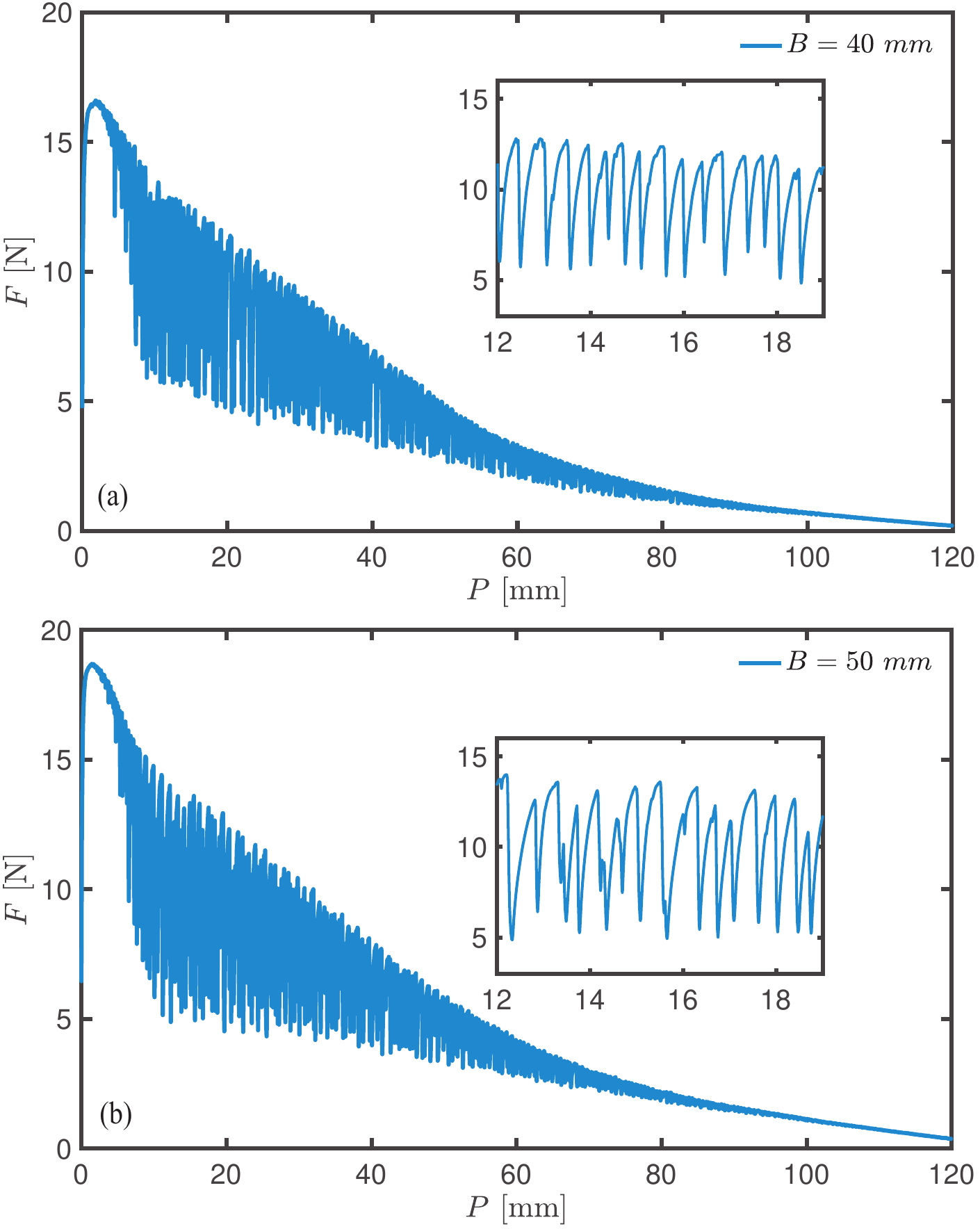}}
\caption{Measuring the average period $\Delta$ between large fluctuations in two uplift tests conducted with $d=$ \SI{0.3}{mm}, $v=$ \SI{5}{mm/min} and two different plate sizes: (a) $B$ = \SI{40}{mm}, (b) $B$ = \SI{50}{mm}. Insets: example of a \SI{7}{mm} displacement window used to count fluctuation events (see text).}
\label{fig:Delta}       
\end{figure}

\subsubsection{A flapping dynamic under the plate}

In contrast with the stick-slip dynamics, drag drops do not seem to be triggered when a force-threshold is reached. Rather, they seem to be triggered when the plate has moved up some distance $\Delta$ since the previous drop (figure \ref{fig:super_zoom}). We propose the physical process illustrated on figure \ref{fig:mechanism}d,e to explain this displacement threshold.

We assume that a conical shape gap is being formed under the plate once it has moved up on a large enough distance. We further consider that the slope angle of the granular packing is bounded by two values, $\alpha^{start}$ and $\alpha^{stop}$, with $\alpha^{start}>\alpha^{stop}$. $\alpha^{start}$ corresponds to an angle of avalanche: it is the maximum  slope angle that the granular surface can sustain statically without collapsing, which would then induce a flow of grains downward.  $\alpha^{stop}$ corresponds to an angle of repose: it is the slope angle at which that flow would stop \cite{al2018review}. 

Our proposed scenario involves a cyclic variation in slope angle between these two extremes angles. Starting from $\alpha^{stop}$, the upward motion of the plate creates a vertical opening at the top of the slope, through which grains would fall. This  contributes to increasing the slope angle by raising the highest point of the slope while the lowest point does not move. Once the critical slope $\alpha^{start}$ is reached, a slope collapse occurs during which grains flow downhill and fill the bottom of the gap. The raise of this point leads to a gradual decrease in slope angle. The avalanche stops when the slope angle reaches $\alpha^{stop}$, and a new cycle begins. 

According to this scenario, the building up of the slope from $\alpha^{stop}$ to $\alpha^{start}$ is driven by the plate's motion. This involves some degree of plasticity as some grains are recirculating from the front to the rear of the plate. In contrast, the dynamics of the avalanche that relaxes the slope from $\alpha^{start}$  back to $\alpha^{stop}$ is independent of the plate's motion. It is a gravity driven flow, which one can compare to a heap-flow \cite{midi2004dense}. This mechanism would be able to relax the drag force even in the limit of an infinitely slow uplift velocity.

This scenario predicts a large drop in drag force when the plate has moved up a distance $\Delta$ given by:

\be \label{eq:Delta}
\Delta = \frac{B}{2} (\tan \alpha^{start} -\tan  \alpha^{stop})
\ee

\noindent This relaxation trigger is purely geometrical and does not depend on the drag force itself.

To quantitatively assess the validity of this model, we have  (i) measured the angle of repose and the angle of avalanche of the glass beads used in the uplift tests and (ii) estimated the distance $\Delta$ in uplift tests conducted with two plate size $B$ = \SI{40} and \SI{50}{mm}. 

The angle of repose was measured by pouring a volume of glass beads on a rough, horizontal surface. $  \alpha^{stop}$ is then defined as the slope angle of the cone once the pouring is stopped. The avalanche angle was measured by subsequently tilting the surface by an angle $\delta \alpha$ until an avalanche occurred on the cone surface. The avalanche angle is then defined as $ \alpha^{start} =  \alpha^{stop}+\delta \alpha$. With this method, we found $\alpha^{stop}=23^\circ$ and $\alpha^{start}=24^\circ$ for glass beads of \SI{300}{\micro\meter}, and $\alpha^{stop}=24^\circ$ and $\alpha^{start}=26.5^\circ$ for glass beads of \SI{50}{\micro\meter}. These values are consistent with values of angle of repose and avalanche reported in \cite{farin2014fundamental}.

The distance $\Delta$ was estimated during uplift tests after the maximum drag force was reached. To exclude smaller amplitude fluctuations, we counted the number of times ($n$) the force dropped by an amount greater or equal to two standard deviations $\sigma$. $\sigma$ was measured as the drag force standard deviation on an interval of $\delta P=$\SI{7}{mm}. $\Delta$ is then given by $\Delta = \delta P/n $. 

Figure \ref{fig:Delta} illustrates the result of this method on two tests conducted with plates of size $B$ = \SI{40} and \SI{50}{mm}, in glass beads of $\SI{300}{\micro\meter}$.  Experimental results point out values of $\Delta$ are \SI{0.43}{mm} and \SI{0.5}{mm} for the plate sizes $B=$ \SI{40}{mm} and \SI{50}{mm}, respectively. This is in agreement with Eq. (\ref{eq:Delta})  which predicts  $\Delta$ = \SI{0.42}{mm} and \SI{0.52}{mm} for these plate sizes. 

Using the same method, experimental results yield $\Delta$ = \SI{1.1}{mm} for a \SI{40}{mm} plate embedded in \SI{50}{\micro\meter} glass beads (see figure \ref{fig:super_zoom}). This is in agreement with Eq. (\ref{eq:Delta}) too,  which predicts  $\Delta$ = \SI{1.07}{mm}.

\subsubsection{Stable and unstable regimes for the drag force}

According to this scenario, a drag instability develops if the plate motion is too slow to feed enough grains to sustain a continuous avalanche. Then, an avalanche can develop and stop before any significant plate's motion, creating a drop in drag force.
  
The time scale at which the plate's motion frees grain at the top of the slope is $t_v = d/v$. The time for the avalanche to develop and stop is difficult to predict quantitatively. Nonetheless, as a gravity-driven flow occurring on a distance proportional to half the plate width, $B/2$, the time for grains to flow down the slope is expected to be proportional to $B/\sqrt{gd}$. This allows us to define a critical plate velocity $v_c$ as:

\be \label{eq:vc}
v_c \approx 2\sqrt{gd}\frac{d}{B}
\ee

\noindent Below this velocity, the avalanche has time to occur and stop before the plate motion can free more grains, and the drag instability develops. Above this velocity, the plate's motion can always free enough grains to sustain a continuous avalanche under the plate, and the drag force is stable. The following two observations support the validity of this scenario.

\begin{figure}[h!]
 {\includegraphics[width=\columnwidth]{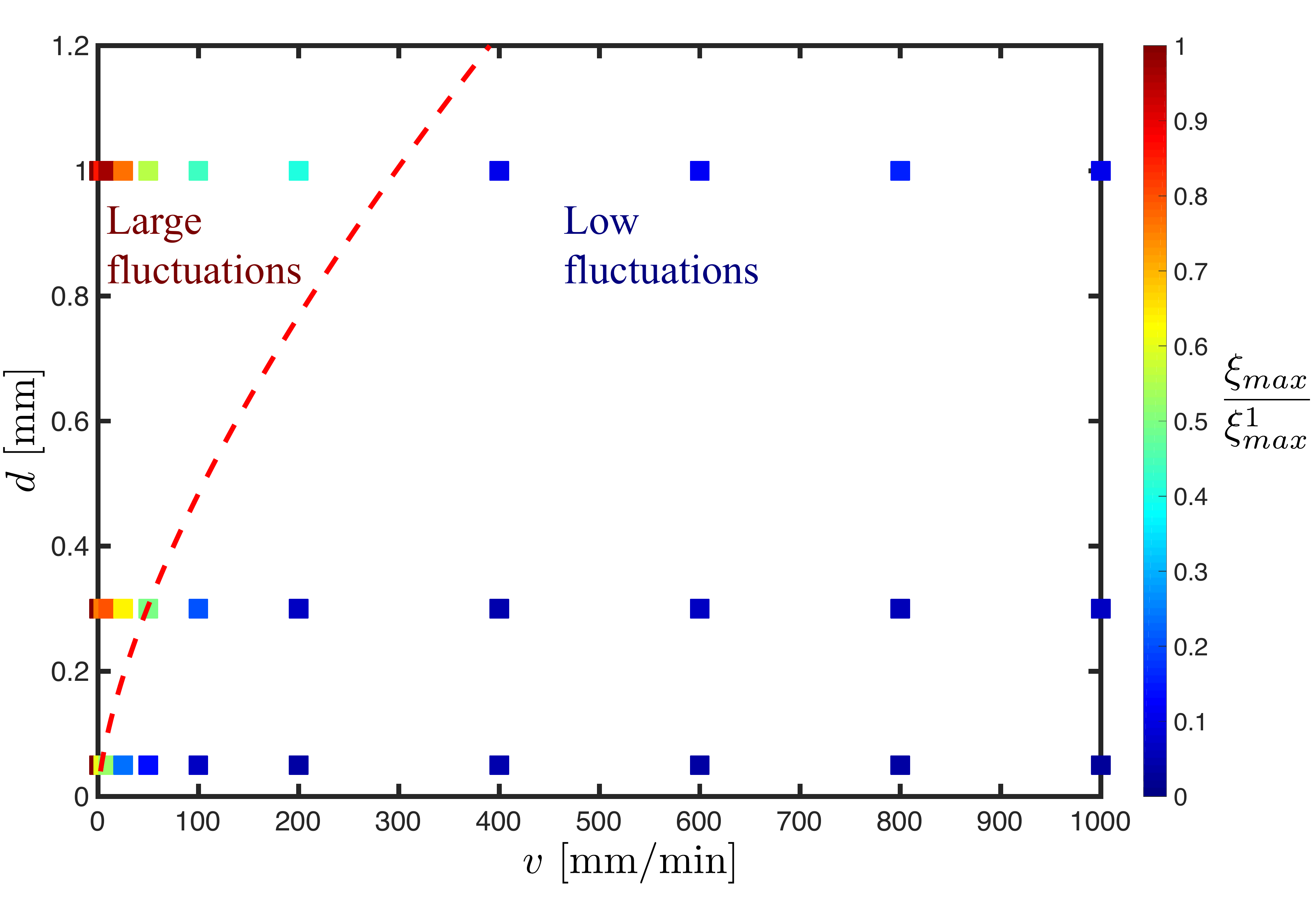}}
\caption{Fluctuation regime map. Each point corresponds to one uplift test performed with a given grain size and a given velocity. Points' color represents the relative magnitude of the drag fluctuations estimated for each test. The dashed line represents Eq. \ref{eq:vc}.}
\label{fig:regime}        
\end{figure}

Firstly, figure \ref{fig:regime} represents a map of the fluctuation magnitude measured with different grain size $d$ and different velocities. The metric used to compare fluctuation magnitudes in any given test is $\xi_{max}/\xi_{max}^1$. $\xi_{max}$ is the maximum fluctuation magnitude - as defined in Eq. (\ref{eq:xi})- measured in during one test. $\xi_{max}^1$ is the maximum fluctuation amplitude measured with the same grain size at the minimum uplift velocity (\SI{1}{mm/min}).

The regime map shows that fluctuation magnitude gradually decreases when the velocity is increased. It further evidences a grain size effect on the typical velocity above which fluctuations vanishes. The prediction of Eq. (\ref{eq:vc}) is qualitatively consistent with this regime map. However, while this model seems capable of delineating the existence of two regimes, it does not provide any prediction for the gradual decrease in fluctuation magnitude as the velocity is increased.

\begin{figure}[h!]
 {\includegraphics[width=\columnwidth]{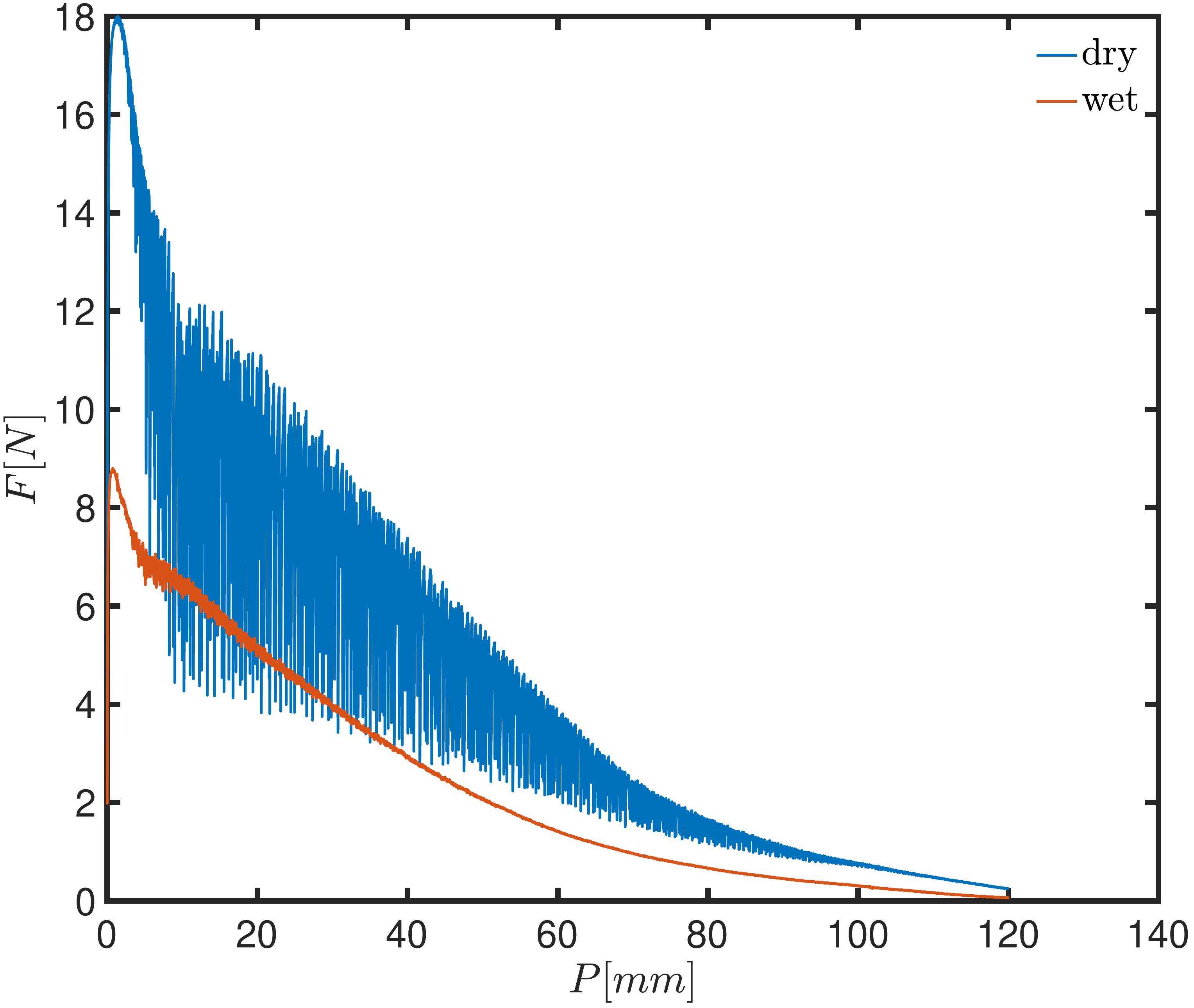}}
\caption{Drag forces measured in dry and fully immersed granular packing with \SI{50}{\micro\meter} grains at $v=$\SI{1}{mm/min}. With water, the drag instability does not develop.}
\label{fig:dry_wet}       
\end{figure}

Secondly, we performed a set of comparative uplift tests in dry and immersed conditions. Both tests were conducted with \SI{50}{\micro\meter} grains at v= \SI{1}{mm/min}; the only difference being the presence or absence of water. The assumption is that the presence of water would slow down the avalanche velocity by inducing viscous forces between grains \cite{cassar2005submarine,rognon2010internal,rognon2011flowing}. According to our proposed scenario, this would reduce the critical velocity $v_c$ for the drag instability to develop. Figure \ref{fig:dry_wet} shows that the drag instability developed in dry conditions but did not with water, which is qualitatively consistent with this prediction. However, it is also possible that the presence of water affects the angle of repose and angle of avalanche of the packing. The absence of fluctuation could also result from  these two angles becoming equal.

\section{Conclusion}

The uplift experiments presented in this paper show that granular drag forces can be unstable at low velocities. This instability is empirically characterised by sudden, quasi-periodic drops in drag force followed by a gradual recovery reminiscent to a non-linear spring behaviour. With our experimental conditions, we measured significant drop amplitudes as large as $60\%$ at the lowest uplift velocity. At high velocities, this instability vanishes and the drag force no longer fluctuates.

The spectral analysis of the drag force fluctuations suggests that an elementary trigger is the displacement of the plate over the size of one grain, with a frequency $d/v$. In contrast, we found that large drops in drag force are controlled by the dynamics of grain recirculation under the plate. In particular, we propose that the drag instability arises from intermittent avalanches under the plate that are triggered every time the plate moves up by a given distance, given by Eq. (\ref{eq:Delta}).  If the typical avalanche duration is longer than the time it takes for the plate to reach this distance, we infer that a continuous avalanche is sustained, which leads to a regime of stable drag force. Accordingly, we propose a simple criterion (Eq. (\ref{eq:vc})) to determine the critical plate velocity below which the drag instability develops and above which it vanishes. 

These results and analysis open several questions. Firstly, our experiments show that the amplitude of large drops gradually decreases with an increasing velocity. However, the models proposed in this paper are not designed for capturing this magnitude, and its underlying mechanisms remain to be established. Secondly, this drag instability was observed here in uplift tests; whether a similar instability develops with different directions of motion, including vertical penetration and horizontal motion also remains to be established. Lately,  a visualisation of the granular flow  around the plate during fluctuations would provide direct support for the proposed mechanisms controlling the instability. We performed quasi-2D tests in this aim, involving a rectangular plate set close to a glass container edge. However, mechanical interactions between the plate and the container made the force measurements in these tests inconclusive. We anticipate that an X-Ray radiography technique such as that introduced in \cite{baker2018x} could enable in-situ visualisation in 3D drag tests.


\section*{Compliance with ethical standards}
Conflict of Interest: The authors declare that they have no conflict of interest.

\section*{Appendix: Testing of the experimental setup}

Two tests were conducted to assess the effect of the shaft and to quantify the stiffness of the loading frame and plate.

\subsection*{Effect of the shaft}
An uplift test was performed without any plate attached to the shaft, as a way to isolate its influence on the drag force.
The shaft was initially embedded at a depth \SI{120}{mm} and then uplifted at a constant velocity of \SI{5}{mm/min}. 
The measured drag force versus displacement response is shown in figure \ref{fig:appendix}a. It evidences a shape similar to the one obtained with a plate. However, the maximum drag force is significantly lower. It is  about \SI{3.2}{N} with the shaft only and \SI{18}{N} with a plate. Also, drag fluctuations did not significantly develop with the shaft only. This confirms that the drag instability is linked to the presence of the plate.

\subsection*{Stiffness of the loading frame/plate system}

\begin{figure}[h!]
 {\includegraphics[width=1\columnwidth]{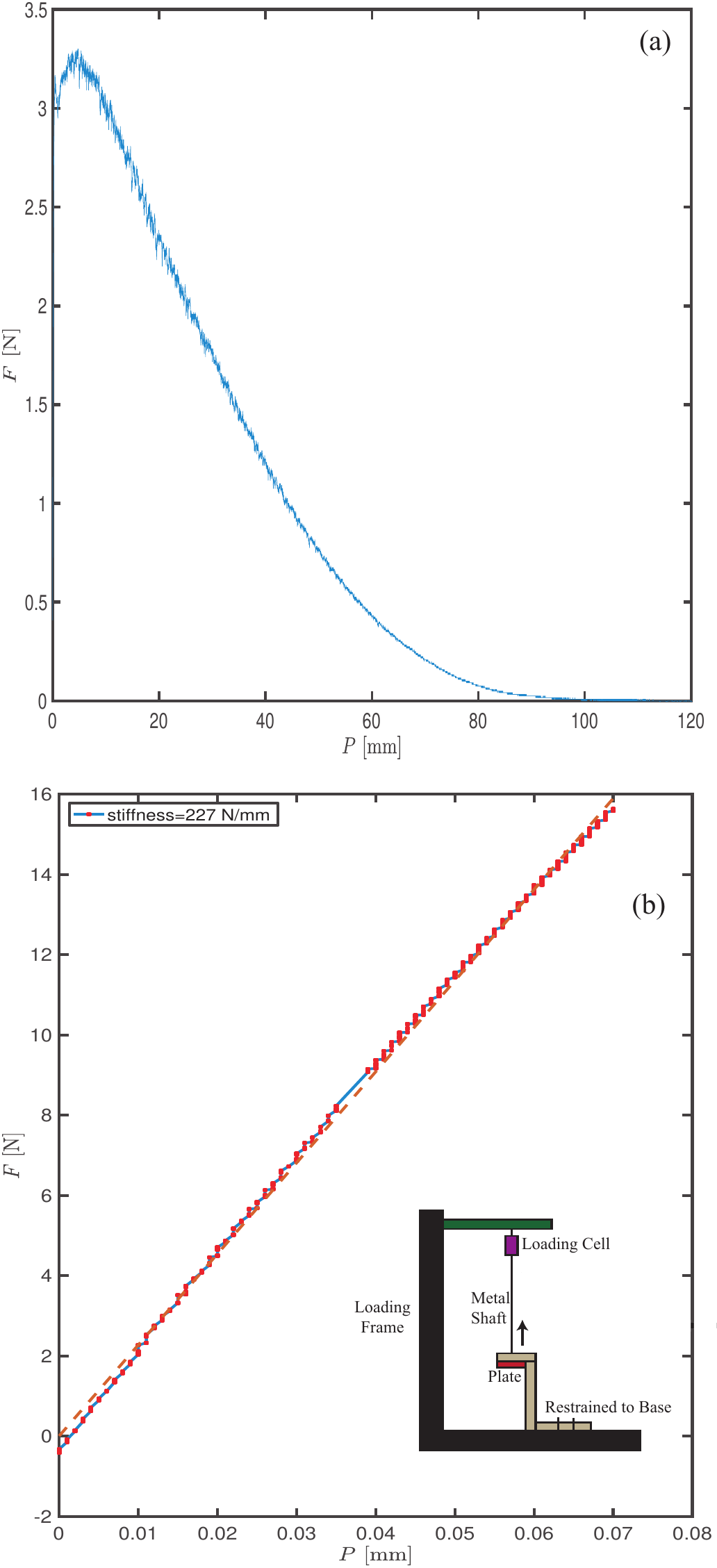}}
\caption{Testing of the uplift experimental device: (a) Drag force measured with a shaft only and no plate (b) Stiffness test performed by holding the plate's edges while applying a vertical force in it centre via the shaft. }

\label{fig:appendix}       
\end{figure}

A specific test was conducted to quantify the effective stiffness $k_0$ of the system comprised of the load cell, shaft and plate. This test involves restraining the edges of the plate to a steel frame fixed to the ground, applying an external force on the middle of the plate via the shaft and measuring the displacement $P$ of the loading frame. This displacement includes elastic deformations of the load cell and shaft and the bending deflection of the plate. Results shown in figure \ref{fig:appendix}b indicate a linear relationship between force and displacement, which we model as: $F \approx k_0 P$. This relationship is valid in the range of forces considered in this study, from zero to \SI{20}{N}. Best fit of the data is obtained for $k_0 = 2.27 \times 10^5$ \SI{}{N/m}.

\end{document}